\documentstyle[12pt]{article}

\catcode`@=11
\def\secteqno{\@addtoreset{equation}{section}%
\def\theequation{\thesection.\arabic{equation}}}
\catcode`@=12
\secteqno
\newcommand{\be}{\begin{equation}}
\newcommand{\ee}{\end{equation}}
\newcommand{\bea}{\begin{eqnarray}}
\newcommand{\eea}{\end{eqnarray}}
\newcommand{\bref}[1]{(\ref{#1})}
\newcommand{\ep}{\epsilon} 
\newcommand{\T}{\theta} 
   
\newcommand{\A}{\alpha} \newcommand{\B}{\beta}
\newcommand{\G}{\Gamma} \newcommand{\D}{\delta}
\newcommand{\gam}{\gamma} 
\newcommand{\K}{\kappa}         
\newcommand{\Lam}{\lambda}
\newcommand{\bLam}{{\bar \lambda}}
          \newcommand{\w}{\omega}

\newcommand{\h}{\eta}

            \newcommand{\s}{\sigma}

\def\pa{\partial}

\def\hM{{\hat{M}}}\def\hN{{\hat{N}}}
\def\CC{{\cal C}} \def\CR{{\cal R}}
\def\CS{{\cal S}} \def\CT{{\cal T}}
\def\CG{{\cal G}} \def\CZ{{\cal Z}}
\def\CL{{\cal L}} 
\def\CA{{\cal A}}
\def\CF{{\cal F}}
\def\CP{{\cal P}}

\newcommand{\nn}{\nonumber}
\def\bd{{\bf{d}}}

\def\ba{\overline}

\def\t{\tilde}
\def\Tt{\tilde \theta}
\def\Ttb{\overline{\tilde \theta}}

\def\l{{\ell}}
\def\bd{{\bf{d}}}
\def\kb{{\bar\kappa}}
\def\bV{{\bf{V}}}
\def\bPi{{\bf{\Pi}}}

\def\ggt{{\Gamma_{11}}}

\newcommand{\slPi}{/ {\hskip-0.27cm{\Pi}}}

\newcommand{\bslPi}{/ {\hskip-0.27cm{\bf{\Pi}}}}
\def\hmu{{\hat{\mu}}}
\def\hnu{\hat{\nu}}

\def\5{\overline}  \def\6{\partial } \def\7{\tilde }
\def\8{\hat }
\newcommand{\ua}{{\underline{{\alpha}}}}
\newcommand{\ub}{{\underline{{\beta}}}}
\newcommand{\uA}{{\underline{{A}}}}

\newcommand{\uz}{{\underline{{z}}}}
\newcommand{\uaa}{{\underline{{a}}}}
\newcommand{\ubb}{{\underline{{b}}}}
\newcommand{\vs}{{\vskip 4mm}}
\def\Dpm{{D($p\hskip-0.75mm-\hskip-1mm1$)}}

\def\limit#1#2{\smash { \mathop{#1} \limits_{#2} }  }

\newcommand{\bZ}{\mbox{\bf ${Z}$}}

\begin{document}

\thispagestyle{empty}
\begin{titlepage}
\vfill
\begin{flushright}
(NUCPHB 8509;revised version 1/June/00)\\
UB-ECM-PF-00/04\\
TOHO-FP-0064
\end{flushright}
\vfill

\begin{center}
{\Large\bf 
 T-duality Covariance of SuperD-branes
}
\par
\vskip 10mm
{\large Kiyoshi\ Kamimura$^{\dagger,1}$ and
Joan Sim\'on$^{\dagger\dagger,2}$,}\par
%\address
{
\medskip
$^{\dagger}$ Department of Physics, Toho University, 
Funabashi\ 274-8510, Japan \\
$^{\dagger\dagger}${\it Departament  ECM, Facultat de F\'{\i}sica\\
Universitat de Barcelona and Institut de F\'{\i}sica d'Altes Energies\\
Av.Diagonal 647, E-08028 Barcelona, Spain
}\par}
\medskip
\end{center}
\vskip 20mm
\begin{abstract}
T-duality realized on SuperD-brane effective actions probing in constant
$G_{mn}$ and $b_{mn}$ backgrounds is studied from a pure world volume
point of view. It is proved that requiring {\em T-duality covariance}
of such actions ``fixes'' the T-duality transformations of the world volume
dynamical fields, and consequently, of the NS-NS and R-R coupling superfields.
The analysis is extended to uncover the mapping of the symmetry structure 
associated with these SuperD-brane actions. In particular, we determine the 
T-duality transformation properties of kappa symmetry and supersymmetry, which
allow us to prove that bosonic supersymmetric world volume solitons of the 
original theory generate, through T-duality, the expected ones in the T-dual
theory. The latter proof is generalized to arbitrary bosonic backgrounds.
We conclude with some comments on extensions of our approach to
arbitrary kappa symmetric backgrounds, non-BPS D-branes and non-abelian
SuperD-branes.
\end{abstract} 
\noindent
{PACS: 11.10.Kk, 11.25.-w \\} 
{Keywords: T-duality covariance, D-branes, kappa symmetry, world volume
solitons}
\par
\vfill
 \hrule width 5.cm
\vskip 2.mm
{\small
\noindent $^1$ E-mail: kamimura@ph.sci.toho-u.jp \\
\noindent $^2$ E-mail: jsimon@ecm.ub.es \\
}

\end{titlepage}
\parskip=7pt

\section{Introduction}
\indent

Many aspects of string physics in ten and eleven dimensions can be
understood from a world volume point of view using D-branes and
M-branes. In particular, duality symmetries of the full string theory
admit a field theory realization on the world volume effective actions
describing the low energy dynamics of these branes. Besides that,
solutions of the classical equations of motion of the latter actions
(world volume solitons) do admit a space time interpretation in terms
of intersections of branes due to the gauge invariant character
of the scalars describing these actions \cite{callan,gary,quim}. 
In this paper we will study
the realization of T-duality on SuperD-brane effective actions probing
in constant $G_{mn}$ and $b_{mn}$ backgrounds, on their symmetry structures
and on supersymmetric bosonic world volume solitons. Given the physical
equivalence between bosonic commutative D-brane gauge theory actions on
such backgrounds and non-commutative gauge theory \cite{sw}, our analysis 
can be seen as a first step towards the supersymmetric extension of such 
an equivalence, giving a full detailed analysis of the commutative side.

It is well known that T-duality admits a field theory
realization in the zero slope limit of closed string theory
giving rise to the
T-duality rules among the NS-NS and R-R massless fields and mapping
$N=2$ $D=10$ IIA Supergravity into $N=2$ $D=10$ IIB Supergravity,
or viceversa \cite{bho}. One can ask whether such a realization
exists in the same limit for the open string sector. This
is answered by studying the T-duality properties of D-brane
effective actions, which describe the low energy dynamics for the
massless open string fields including their interactions with the
massless closed string sector \cite{dbi,Pol}. In \cite{old} it
was proved that the double dimensional reduction of a D$p$-brane
action yields the direct dimensional reduction of a \Dpm-brane.
Their approach was based on the already known T-duality rules
mapping type IIA/IIB backgrounds derived in \cite{bho}. It was
later proved in \cite{joan} that the latter set of transformations
could be derived from a pure world volume perspective, by requiring
{\em T-duality covariance} of the Dirac-Born-Infeld (DBI) and the
Wess-Zumino (WZ) terms appearing in the D-brane effective action.

{\em  T-duality covariance} is the most natural requirement having in mind
the conformal field theory description of D-branes in terms of
open strings \cite{pol1}. D$p$-branes appear as hyperplanes
on which open strings can end. The dimensionality $(p+1)$ depends on the
number of scalar fields satisfying Neumann boundary conditions (b.c.).
Since under a longitudinal T-duality,\footnote{By a longitudinal T-duality,
we mean a T-duality along a direction parallel to the hyperplane defined
by the initial D$p$-brane.} a Neumann b.c. is transformed
into a Dirichlet b.c., we are left with an open string whose end points
are constrained to move in a $p$-dimensional hyperplane {\it i.e.}
\Dpm-brane. Although the number of bosonic massless states in the open
string spectrum remains invariant $({\bf\underline 8})$, the number of bosonic 
scalar ones increases by one, while the number of bosonic vectorial ones
decreases by the same amount. In other words, while the original
bosonic massless open string spectrum fits into a vector supermultiplet
in $(1,p)$ dimensions, the T-dual one fits into a vector supermultiplet
in $(1,p-1)$ dimensions \cite{str}. Thus, any effective field theory
description of the initial and T-dual open string sectors should be
a field theory realization of such vector multiplets in the corresponding
dimensions. Since both of them are known to be of DBI + WZ type form
\cite{dbi,Pol}, the requirement of {\em T-duality covariance} is certainly
justified.

The analysis done in \cite{old,joan} shows that the right way to realize
a longitudinal T-duality on D-brane effective actions is to apply a double
dimensional reduction, which requires the existence of an isometric
direction, but without rewriting the ten dimensional background fields
in terms of the nine dimensional ones \footnote{It would also be interesting
to study transverse T-duality on abelian D-brane effective actions,
generalizing the approach followed in Matrix theory compactifications
\cite{aconnes}, but this is beyond the scope of the present paper.}. Such a 
reduction consists of a partial gauge fixing of the world volume 
diffeomorphisms to fix in which direction the original D-brane is wrapping,
and a functional truncation that discards the non-zero modes of the 
dynamical fields in the infinitely massive ($R\to 0$) limit.
This is again consistent with the conformal field theory picture,
because whenever the radius $R$ of the circle along which we are T-dualizing
becomes much more smaller than the string scale $\sqrt{\alpha'}$,
physics have a much more natural description in terms of the T-dual theory,
which in our case is a $p$ dimensional field theory; the T-dual \Dpm-brane
effective action.

The extension of the analysis done in \cite{joan} to the supersymmetric
case is conceptually straightforward. When describing superD-branes, one must
also include fermionic scalar fields 
$\theta_i$  $(i=1,2)$ having different ten dimensional chiralities 
in type IIA, and 
$\theta'_i$  with the same chiralities in type IIB.
Being world volume scalars, we will just keep their zero modes along
the direction of dualization.
We will show how the requirement of {\em T-duality covariance} fixes the
necessary chirality changing mapping between
the fermionic degrees of freedom describing  type IIA/IIB D-branes
in addition to the one for bosonic fields. Furthermore, this mapping of
dynamical degrees of freedom indeed maps the original DBI and WZ 
terms into the T-dual ones, thus generalizing not only previous bosonic 
analysis but also the supersymmetric one \cite{schw} in which kappa gauge 
symmetry was fixed.

Our proof is not only concerned with effective actions but also uncovers
their gauge and global symmetry structures, thus 
generalizing the corresponding bosonic analysis done in \cite{joan}.
In particular, we will show how kappa symmetry and supersymmetry
transformations of D-branes probing in constant $G_{mn}$ and $b_{mn}$ 
are mapped under T-duality. 
Since we will always be concerned with T-duality 
performed along an isometric direction of the background 
%whose Killing spinors are independent of the coordinate 
%adapted to the isometry, 
it ensures the preservation of supersymmetry under the dualization 
\cite{ert}. In this way, we will find out the T-duality 
transformation properties of the $\G_\kappa$ matrix appearing in kappa 
symmetry transformations.

We shall also study the effect of T-duality on bosonic supersymmetric
world volume solitons. For any super-brane action in any background
compatible with kappa symmetry, such configurations must satisfy
\begin{equation}
\G_{\kappa} \ep = \ep
\label{spc0}
\end{equation}
which is from now on called kappa symmetry preserving condition \cite{ertg}.
Here $\ep$ is a linear combination of Killing spinors of the
background and the number of supersymmetries preserved by the combined
background/brane configuration is the number of linearly-independent
solutions of (\ref{spc0}).
Such an equation involves, generically,
a set of constraints among the excited dynamical fields or {\em BPS equations}
and a set of supersymmetry projection conditions $\CP_i\ep = \pm \ep$
determining the type of branes described by the configuration.
We will argue that the functionally truncated and partially
gauge fixed BPS equations are the corresponding BPS equations describing
the {\em supersymmetric T-dual configuration}.
The supersymmetric projection conditions $\CP'_i\ep' = \pm \ep'$ are
obtained from the initial ones $\CP_i\ep = \pm \ep$
by rewriting them in terms of the
T-dual Killing spinors $\ep=\ep(\ep')$. This mapping having the same
form as the one among dynamical fermionic fields $\theta=\theta(\theta')$
which would have already been fixed by {\em T-duality covariance} of the 
SuperD-brane effective action.

It is interesting to remark that such mapping of supersymmetric world 
volume solitons we have described is nothing but a world volume
realization of a well known algebraic mapping. D-brane effective
actions are supersymmetric field theories, being ten dimensional
target space covariant, whose group of global isometries contains
the isometry supergroup of the background \cite{paul}. In our case,
they provide a field theory realization of $N=2$ $D=10$
IIA(IIB) SuperPoincare algebras for $p$ even(odd), which are known
to be related by some transformation of their generators, reminiscent
of T-duality \cite{paul1}. Given such a relation, BPS states in string
theory admit different realizations. One is purely algebraic and is based
on the saturation of the BPS bound in the supersymmetry algebra. Such a
bound is exactly the same one derived from a hamiltonian analysis
of brane effective actions \cite{quim}, giving rise to some set of BPS 
equations, this being the field theoretical description of such states. 
These BPS equations derived from the phase space formulation of D-branes
are entirely equivalent to the resolution of (\ref{spc0}) due to the
connection between the supersymmetry algebra and the structure of the 
kappa symmetry projector \cite{man}.

The mapping of BPS equations and supersymmetry projection conditions
will be illustrated by some examples. To begin with, we will study
the effect of T-duality on BIon and dyon solutions of D-brane effective
actions probing SuperPoincar\'e $(b_{mn}=0)$ background. BIons are mapped
among themselves, in agreement with the conformal field theory picture,
while dyons are mapped to a non-threshold bound state of a D2-brane
and a fundamental string parallel to it intersecting in a point with a
D2-brane, altogether giving a $\nu={1\over 4}$ threshold bound state.
Later, we concentrate on solitons in $b_{mn}\neq 0$ constant backgrounds.
In particular, we will study T-duality on tilted dyons and tilted
BIons on non-threshold bound states of D-strings and D3-branes. We will
show that, generically, just as constant flux of magnetic field on the
D-brane is seen as D-branes at angles in the T-dual picture, constant
electric field (induced by the electric components $b_{0a}$) boosts the
configuration in the direction along which we are T-dualizing.

Having proved the mapping of supersymmetric world volume solitons under
T-duality for constant $G_{mn}$ and $b_{mn}$ backgrounds 
we extend the proof for an arbitrary bosonic background, relying on the
$\theta=0$ condition characterizing any bosonic configuration and the
standard T-duality rules mapping bosonic backgrounds, from which we
can derive the T-duality transformation properties of the bosonic kappa
matrix $\G_\kappa|_{\theta=0}$. In this way, we show the generating
solution character of T-duality transformations in the low energy
description of the open string sector, in close analogy with such generating
character already known in type IIA/IIB supergravities describing
the massless closed string sector.

In sections 2-4 we discuss the T-duality covariance of the D-brane actions
and their symmetry properties. Some details are given in appendices.
In section 5 the mapping of world volume solitons is discussed with
some examples. The extension to arbitrary bosonic background is examined in 
section 6. 
In the last section we comment on possible generalizations and/or extensions
of our present work. They include T-duality of D-brane effective actions
for arbitrary kappa symmetric backgrounds, non-BPS D-branes and non-abelian
SuperD-branes.

%%%%%%%%%%%%%%%%%%%%%%%%%%%%%%%%%%%%%%%%
\section{Effective action and symmetry structure}
\indent

The effective Lagrangian density of a type IIA D$p$-brane is a sum of 
DBI and WZ terms \cite{BT,Cw,Shgf}
\bea
\CL&=&\CL^{DBI}~+~\CL^{WZ},
\label{totlag}\\
\CL^{DBI}&=&-~T_p~\sqrt{-det(\CG_{\mu\nu}+\CF_{\mu\nu})},
\label{DBI}\\
\CL^{WZ}~&=&~[L^{WZ}]_{p+1},~~~~~L^{WZ}~=~-~T_{p}~\CC~e^{\cal F},
\label{WZ}
\eea
where $T_{p}$ is the D$p$-brane tension scaling as     
$T_{p}\propto ({g_s\alpha^{\prime (p+1)/2}})^{-1}$. 
Due to the fact that we are considering constant backgrounds
all the dependence of the constant
dilaton background is included in the string coupling constant through
$~g_s=e^{\phi_0}$.

The DBI term depends on the world volume induced metric
\bea
\CG_{\mu\nu}&=&E_\mu^{~~\uaa}~E_\nu^{~~\ubb}~\h_{\uaa\ubb}\,,
\eea
where $E_\mu^{~~\uaa}$ stand for the components of the supersymmetric
invariant one forms
\bea
E^{\uA}&\equiv&dZ^M~e_M^{~~\uA}~=~d\sigma^\mu~\pa_\mu Z^M~e_M^{~\uA}~\equiv~
d\sigma^\mu~E_\mu^{~\uA},
\label{oneform}\eea
which for the backgrounds considered in this paper take the form
\bea
E^{\uaa}&=&d\7x^{\uaa}~+~\ba{\7\T}\G^{\uaa}d{\7\T}~\equiv~\Pi^{\uaa},~~~~~~~
\7x^{\uaa}~\equiv~x^me_m^{~~\uaa},
\label{susyinv} \\
E^{\ua}&=&d{\7\T}^{\ua},~~~~~~~~~~~~~~~~~~~~~~~~~~~~~~
{\7\T}^{\ua}~\equiv~\T^\A e_\A^{~\ua}.
\label{susyinv1}
\eea
$e_m^{~~\uaa}$ and $ e_\A^{~\ua}$ are constant components of the 
supervielbeins and $Z^M~\equiv~(x^m,\T^\A)$ parametrize the target
superspace. It also depends on the supersymmetric invariant
\bea
\CF~=~dV~-~B\,,
\label{susyiF}
\eea
where $B$ stands for the NS-NS two form, containing additional constant 
bosonic components ($b_{mn}$)
\bea
B&=&\frac12~dZ^MdZ^N~B_{MN}~=~
-\ba{\7\T}\G_{11}\G_\uaa d\t\T(d\t x^\uaa +\frac12\ba{\7\T}
\G^\uaa d\t\T)~
+~\frac12~dx^mdx^n~b_{mn},~~~~~
\label{twoform}
\eea 
but still satisfying the supergravity constraint
\bea
H~=~dB&=& -E^\ua(C\G_{11}\G_\uaa)_{\ua\ub} E^{\ub}~E^\uaa~\equiv~
-(\ba E~\G_{11}\slPi~E),
\eea
where $\slPi=\G_\uaa \Pi^\uaa$. For type IIB D$(p-1)$-branes, the dynamical
fields will be indicated by ``primes'' and $\Gamma_{11}$ must be replaced
by $\tau_3$.

\vs

Concerning the WZ term in \bref{WZ}, it is the
$p+1$ form part of a symbolic sum of differential forms $L^{WZ}$
\cite{old,Doug} satisfying
\bea
d L^{WZ}~&=&~-~T_{p}~\CR~e^{\cal F},~~~~~~~~
\label{RRIIA}
\eea
where $\CR$ is the field strength of the R-R gauge potential $\CC$.
The R-R field strength $\CR$ is  expressed in type IIA as
\bea
\CR&=&\ba E~\CC_A(\slPi)~E,~~~~~~~~~~~{\cal C}_A(\slPi)~=~\sum_{\l=0}~
(\Gamma_{11})^{\l+1}~\frac{\slPi^{2\l}}{(2\l)!}
\eea
whereas in type IIB as
\bea
\CR'&=&-\ba E'~\CS_B(\slPi')~\tau_1~E',~~~~~~~~~~~\CS_B(\slPi')~=~\sum_{\l=0}
(\tau_{3})^{\l}~\frac{\slPi^{'2\l+1}}{(2\l+1)!}.
\eea

Denoting the set of fields described by the SuperD-brane effective
action \bref{totlag} by 
\begin{equation}
\{\phi^i\}=\{Z^M,V_\mu\}\, ,
\end{equation}
we will decompose the infinitesimal transformations $(\7s\phi^i)$ 
leaving the effective action invariant, into gauge $(s\phi^i)$ and 
global $(\Delta\phi^i)$ ones. The set of gauge symmetries involves 
world volume diffeomorphisms $(\xi^\mu)$, an abelian $U(1)$ gauge symmetry
$(c)$ and kappa symmetry $(\K)$. They are given by
\begin{eqnarray}
s\7 x^\uaa& = & \xi^\mu\6_\mu \7x^\uaa + \delta_\K \7x^\uaa 
= \xi^\mu\6_\mu \7x^\uaa - 
\delta_{\kappa}\5{\7\theta} \Gamma^\uaa \7{\theta} \label{gx}, \\
s{\7\theta}^\ua & = & \xi^\mu\6_\mu{\7\theta}^\ua + 
\delta_{\kappa}{\7\theta}^\ua \label{gt}, \\
sV_\mu & = & \xi^\nu\6_\nu V_\mu + V_\nu \6_\mu \xi^\nu + \6_\mu c 
+ \delta_\K V_\mu ,
\end{eqnarray}
where the kappa symmetry transformation for the gauge field $\delta_\kappa
V_\mu$ is determined by requiring the invariance of the gauge
invariant tensor $\CF$ in \bref{susyiF} as
\bea
\delta_\K V_\mu =-\delta_\kappa \5{\7\theta}\Gamma_{11}
\Gamma_\uaa\7\T\left(\6_\mu \7x^\uaa - {1\over 2}\5{\7\theta}
\Gamma^\uaa \6_\mu {\7\theta}\right) 
 +{1\over 2}\delta_\kappa\5{\7\theta}\Gamma^\uaa{\7\theta}~
\5{\7\theta}\Gamma_{11}\Gamma_\uaa\6_\mu {\7\theta}
+\D_{\kappa} x^m\6_\mu x^nb_{mn} \,, 
\label{gv}
\eea
while $\delta_\kappa \7\theta$ is fully determined by
\bea
\delta_{\kappa}\ba{{\7\theta}}&=&\kb(1-\gam^{(p)}),~~~~~~~
\gam^{(p)}=\frac{\rho^{(p)}}{\sqrt{-det(\CG+\CF)}}.
\label{deltatb}
\eea
$\rho^{(p)}$ is the $(p+1)$ world volume form coefficient of 
$\CS_A(\slPi)e^\CF$ for type IIA theory,
\bea
\rho^{(p)}&=&[\CS_A(\slPi)e^\CF]_{p+1},~~~~~~~~
 \CS_A(\slPi)~=~\sum_{\l=0}(\ggt)^{\l+1}\frac{\slPi^{2\l+1}}{(2\l+1)!}
\eea
while for type IIB \Dpm-brane  
\bea
\rho^{(p-1)}&=&-[\CC_B(\slPi')e^{\CF'}\tau_1]_{p},~~~~~~~~
\CC_B(\slPi')~=~\sum_{\l=0}(\tau_3)^{\l+1}\frac{\slPi^{'2\l}}{(2\l)!}.
\eea
The set of global symmetries involves supersymmetry $(\ep)$, bosonic
translations $({\rm a}^m)$ and Lorentz transformations 
$(\omega^{mn})$.\footnote{Along the whole paper, we will not take into
account the infinite number of non-trivial global symmetries existing
for the D-string and D0-brane effective actions \cite{ours}, even
though our conclusions also apply to them.}
They act as follows :
\bea
\Delta \7x^{\uaa} &=& \delta_\ep\7x^{\uaa} + \delta_{\rm a} \7x^{\uaa} +
\delta_\omega \7x^{\uaa} =  \5{\7\epsilon}\Gamma^\uaa \7\theta +
{\rm a}^\uaa+\w^\uaa_{~~\ubb}\7x^\ubb \label{glx} ,\\
\Delta {\7\theta}^\ua &=& \delta_\ep {\7\theta}^\ua + \delta_\omega
{\7\theta}^\ua = \7\epsilon^\ua + {1\over 4}\omega^{\uaa\ubb}\left(
\G_{\uaa\ubb}\7\T\right)^\ua \label{glt}, \\
\Delta V_\mu &=& \delta_\ep V_{\mu} + \delta_{\rm a} V_\mu + 
\delta_\omega V_\mu
\nn\\
&=& \5{\7\epsilon}\Gamma_{11}\Gamma_\uaa {\7\theta}\,
\6_\mu \7x^\uaa - 
{1\over 6}\left(\5{\7\epsilon}\Gamma_{11}\Gamma_\uaa{\7\theta}~
\5{\7\theta}\Gamma^\uaa \6_\mu{\7\theta} + 
\5{\7\epsilon}\Gamma_\uaa {\7\theta}~
\5{\7\theta}\Gamma_{11}\Gamma^\uaa \6_\mu{\7\theta}\right) \nn \\
& & +\Delta x^{m} \6_\mu x^n~b_{mn}.
\eea

%%%%%%%%%%%%%%%%%%%%%%%%%%%%%%%%%%%%%%%%
\section{T-duality covariance of SuperD-branes}
\indent

In this section, we will find the constraints derived from the T-duality
covariance requirement on the DBI term (\ref{DBI}) %, when it is 
realizing
a longitudinal T-duality transformation $(\CT_{\parallel})$
on SuperD-brane effective actions, whose solution is given in the appendix.
Afterwards, it will be proved that such a solution maps the WZ terms
(\ref{WZ}) of both theories.

It was argued in the introduction that $\CT_{\parallel}$
was conveniently realized on the world volume action as a kind of
double dimensional reduction. The latter consists of a partial gauge
fixing of the world volume diffeomorphisms
\begin{equation}
z=\rho, ~~~~~~~~~~~~~~~~~~x^m\equiv
\{x^{\8m},z \},~~\s^\mu\equiv\{\s^\hmu,\rho \},
\label{pardif}
\end{equation}
saying in which direction the D-brane is locally wrapping the circle
of radius $R$, besides a functional truncation
\begin{equation}
\6_\rho \phi^{{\8{\i}}}=0 \quad , \quad 
\{\phi^{{\8{\i}}}\}=\{x^{\8m},\theta^\A,V_\mu\}
\end{equation}
that discards all but the zero modes of the rest of dynamical fields
in the limit $R\longrightarrow 0$.

As in the analysis of degrees of freedom done in \cite{old,joan},
we will allow the following relations among $\{\phi^{{\8{\i}}}\}$ and
$\{\phi^{'{\8{\i}}}\}$
\bea
V_\rho&=&\h~{z'},~~~~~~V_\hmu~=~V_\hmu', \\
Z^{'\hM'}&=&(x^{'\8m},~\T_i^{'\A})~=~Z^\hN~\G_{\hN}^{~~\hM'}~,~~~~~~~~~~~~
Z^\hM~=~(x^{\8m},~\T_i^\A),
\label{cons}
\eea
where $\h$ and $\G_{\hN}^{~~\hM'}$ are some set of constants.

Requiring T-duality covariance for the DBI action (\ref{DBI}), 
\begin{equation} 
-T_p\,\int d^{p+1}\sigma \,\sqrt{-det\left(\CG+\CF\right)}
\limit{\longrightarrow}{\CT_{\parallel}}
-T'_{p-1}\,\int d^{p}\sigma \,\sqrt{-det\left(\CG'+\CF'\right)}
\end{equation}
under the assumptions (\ref{pardif}-\ref{cons}), allow us to derive
\footnote{The derivation is left to Appendix \ref{appA}.}
a relation between brane tensions
\begin{equation}
T'_{p-1}= T_p\,R,~~~~~~~~~~
R\equiv\int d\rho \sqrt{\CG_{\rho\rho}}
\label{ttension}
\end{equation}
and a set of constraints among the background superfields
\bea
\frac{\h^2}{G_{zz}}&=&{G'_{zz}},
\label{eq11}
\\
{{-\h}\over G_{zz}}~G_{\hM z}&=&\G_\hM^{~\hM'}~B'_{\hM' z},
\label{eq41}
\\
{{-\h}\over G_{zz}}~B_{\hM z}&=&\G_\hM^{~\hM'}~G'_{\hM' z},
\label{eq21}
\\
G_{\hM\hN}-{1\over {G_{zz}}}
\{ G_{\hM z}G_{\hN z}-B_{\hM z}B_{\hN z}\} &=&(-)^{(M+M')N}
\G_{\hM}^{~\hM'}\G_{\hN}^{~\hN'}
G'_{\hM'\hN'},
\label{eq31}
\\
B_{\hM\hN}-{1\over {G_{zz}}}(-)^{MN}
\{ B_{\hM z}G_{\hN z}-G_{\hM z}B_{\hN z}\}&=&(-)^{(M+M')N}
\G_{\hM}^{~\hM'}\G_{\hN}^{~\hN'}
B'_{\hM'\hN'}.
\label{eq51}
\eea
Equation (\ref{ttension}) is consistent with the T-dual tension,
$T'_{p-1} \propto ({g'_s\alpha^{\prime p/2}})^{-1}$ since
$g'_s = \frac{g_s\sqrt{\alpha'}}{R}$. The latter is equivalent to the
standard T-duality transformation for the dilaton background field,
\bea
\phi^\prime=\phi -\frac{1}{2}\log\vert G_{zz}\vert.
\eea
Equations (\ref{eq11})-(\ref{eq51}) can be interpreted as the T-duality rules
for the NS-NS background superfields considered in this paper. 
%Due to the fact that 
Since we already know such superfields, equations 
(\ref{eq11})-(\ref{eq51}) can be used to fix the set of constants introduced
in (\ref{cons}). The analysis is carried out in appendix \ref{appA} to which
we refer for further details. It is nevertheless natural to expect two 
different sets of constraints, due to the fact that these superfields admit 
an expansion in the fermionic variables $\theta$. The first set has to do
with the bosonic components of such superfields. They are the usual
T-duality rules for the bosonic NS-NS background fields expressed in terms
of the vielbeins
\begin{eqnarray}
\frac{b_{\8mz}}{\Lam}&=&\G_{\8m}^{~\8n}~e'_{\8n\uz},
\label{eqbed}\\
e_{\8m\uz}&=&\G_{\8m}^{~\8n}~\frac{b'_{\8nz}}{\Lam'},
\label{eqebd}\\
e_{\8m}^{~~\8\uaa}&=&\G^{~~\8m'}_{\8m}{e'_{\8m'}}^{\8\uaa},
\label{eerel}\\
b_{\8m\8n}&=&
\G^{~~\8m'}_{\8m}~\G^{~~\8n'}_{\8n}~[~b'_{\8m'\8n'}~-~
\frac{b'_{[\8m'z}}{\Lam'}~
e'_{\8n']\uz}],
\label{eqbmn}
\end{eqnarray}
where we have splitted the flat tangent space indices as
$\uaa=(\8\uaa,\uz)$. 
We have also made use of a local $SO(1,9)$ rotation in both type IIA/B
tangent spaces to choose
\bea
e_z^{~~\uaa}&=&\Lam~\D_{\uz}^{~~\uaa},~~~~~~
e_z^{'~~\uaa}~=~\Lam'~\D_{\uz}^{~~\uaa}
\eea
with $\Lam$ and $\Lam'$ being some constants standing for $\sqrt{G_{zz}}$ and
$\sqrt{G_{z'z'}}$, respectively. 

The second set of constraints, related to the fermionic components
of the NS-NS superfields, fixes the chirality change mapping
of the target space spinor fields up to signs
\bea
\7\T_1&=&a_2~\7\T'_2,~~~~~~
\7\T_2~=~-a_1~\G_\uz~\7\T'_1,
\label{tch2}
\eea
where 
\bea
\G_{11}\7\T_i&=&(-1)^{i+1}\7\T_i,~~~~~~\tau_{3}\7\T'_i~=~(-1)^{i+1}\7\T'_i,
\eea
and ${a_1}^2={a_2}^2=1$. From these equations and the algebra of Pauli
matrices, the following set of identities can be derived
\begin{eqnarray}
& \5{\7\theta}\Gamma^{\8\uaa} \6_{\hmu}{\7\theta} = 
\5{\7\theta}'\Gamma^{\8\uaa}\6_{\hmu}{\7\theta}'
\quad , \quad 
\5{\7\theta}\Gamma_{11}\Gamma^\uz \6_{\hmu}{\7\theta} = 
-\5{\7\theta}'\Gamma^{\uz}\6_{\hmu}
{\7\theta}', & \nonumber \\
& \5{\7\theta}\Gamma_{11}\Gamma_{\8\uaa} \6_{\hmu}{\7\theta} = 
\5{\7\theta}'\tau_3\Gamma_{\8\uaa}
\6_{\hmu}{\7\theta}' \quad , \quad
\5{\7\theta}\Gamma^\uz \6_{\hmu}{\7\theta} = -
\5{\7\theta}'\tau_3\Gamma^\uz \6_{\hmu}
{\7\theta}'. &
\label{conditions}
\end{eqnarray}
\noindent 
 The same form of formulas are also applied when 
$\T~(\T')$ and $\pa\T~(\pa\T')$ are replaced by any IIA(IIB) spinors
related by \bref{tch2}.

The T-duality covariance of the WZ actions follows from \bref{dWZA}
in Appendix B,
\bea
(\CR e^\CF)~=~-{a_1a_2~\Lam}~(\CR' e^{\CF'})~d\rho.
\eea
The WZ action of IIB D-brane is obtained by integrating the IIA one over 
$\rho$
if $~a_1a_2=-1~$,
\bea 
\CL_A^{WZ} &
\limit{\longrightarrow}{\CT_{\parallel}} &
\CL_B^{WZ}.
\eea

%%%%%%%%%%%%%%%%%%%%%%%%%%%%%%%%%%%%%%%%%%%%%%%%%%%%%%%%%%%%%%%
\section{T-duality and symmetry structure}
\indent

Let us define by ${\cal A}$ the subspace of the field configuration
space defined by the partial gauge fixing $(z=\rho)$ and functional
truncation $(\6_\rho \phi^{{\8{\i}}}=0)$. In general, ${\cal A}$ is not
left invariant under $\7s\phi^i$, so we must require two consistency
conditions. The first one ensures that $\7sz$ will not move our
configuration from the gauge slice defined by the partial gauge
fixing,
\begin{equation}
\7sz|_{{\cal A}}=0~~~~ \Longrightarrow~~~~ 
\xi^\rho=-\left(\delta_\K z + \Delta z
\right)|_{{\cal A}}.
\label{comp}
\end{equation}
The second consistency condition ensures that $\7s\phi^{{\8{\i}}}$ will
respect the functional truncation
\begin{equation}
(\6_\rho \7s\phi^{{\8{\i}}})|_{{\cal A}}=0\, ,
\label{trunc}
\end{equation} 
constraining the gauge and global parameters. A sufficient solution of this 
set of constraints is given by
\bea
& \xi^{\hmu} = \xi^{\hmu}(\sigma^{\hnu}) \quad , \quad
\K = \K(\sigma^{\hnu}) \quad , \quad c(\sigma^{\hmu},\rho)=
\8c(\sigma^{\hmu}) 
+ A\rho, & \\
\nn\\
& \omega^{\underline{z}\8\uaa}=0, & \label{constraint}
\eea
which is explicitly breaking the global symmetry group $SO(1,9)$ into
the $SO(1,8)$.  
The fixation of the diffeomorphism with respect to $\rho$ 
will induce compensating transformations through \bref{comp}
modifying the transformation property of $V_{\hmu}$.
In this way, the symmetry structure $(\7s\phi^{\8{\i}}|_
{{\cal A}})$ of the partially gauge fixed truncated action is found. 
The gauge symmetries are given by \footnote{It should be understood that the 
transformations appearing in the right hand side of the forthcoming equations
must be computed in ${{\cal A}}$.}
\bea
s \7x^{\8\uaa}|_\CA &=& \xi^{\hmu} \6_{\hmu} \7x^{\8\uaa}  -
\delta_\kappa\5{\7\theta}\Gamma^{\8\uaa} {\7\theta} \label{gf1} \\
s {\7\theta}^\ua|_\CA & = & \xi^{\hmu} \6_{\hmu} {\7\theta}^\ua +
\delta_\kappa {\7\theta}^\ua \label{gf2} \\
s V_\rho|_\CA &=& \xi^{\hmu}\6_{\hmu}V_\rho -
\delta_\kappa\5{\7\theta} \Gamma_{11}\Gamma_\uz {\7\theta}~\Lam +
\delta_\kappa x^{\8m} b_{\8m z} \label{gf3} \\
s V_{\hmu}|_\CA &=& \xi^{\hnu}\6_{\hnu} V_{\hmu} + V_{\hnu} \6_{\hmu} 
\xi^{\hnu} + \6_{\hmu}c^* + \delta^*_\kappa V_{\hmu}
\label{gf4}
\eea
and the global symmetries are 
\bea
\Delta \7x^{\8{\uaa}}|_{{\cal A}} &=& \delta_\ep\7x^{\8{\uaa}} + 
\delta_{\rm a} \7x^{\8{\uaa}} + \delta_\omega \7x^{\8{\uaa}} =  
\5{\7\epsilon}\Gamma^{\8{\uaa}} \7\theta +
{\rm a}^{\8{\uaa}}+\w^{\8{\uaa}}_{~~\8{\ubb}}\7x^{\8{\ubb}} \label{glgfx} \\
\Delta {\7\theta}^\ua|_{{\cal A}} &=& \delta_\ep {\7\theta}^\ua + \delta_\omega
{\7\theta}^\ua = \7\epsilon^\ua + {1\over 4}\omega^{\8{\uaa}\8{\ubb}}\left(
\G_{\8{\uaa}\8{\ubb}}\7\T\right)^\ua \label{glgft} \\
\Delta V_\rho|_\CA &=& A + \5{\7\epsilon}\Gamma_{11}\Gamma_\uz {\7\theta}~\Lam
+ \Delta x^{\8m} b_{\8m z} \label{gfg1} \\
\Delta V_{\hmu}|_\CA &=& \delta^*_\ep V_{\hmu} + \delta^*_{\rm a} V_{\hmu} +
\delta^*_\omega V_{\hmu},
\label{gfg2}
\eea 
where 
\bea
\delta^{*}_i V_{\hmu} = \delta_i V_{\hmu}+\delta_i z~\6_{\hmu} V_\rho,~~~~~~~~~
i=(\K,\ep,{\rm a},\w)
\eea
and
\bea
c^*&=&c+ V_\rho\xi^\rho.
\eea

In the following we will see that these transformations give the right 
transformation properties of the T-dual variables, that is we will prove
that the whole symmetry structure of these theories is properly mapped
under T-duality. To begin with, $(p-1)$ dimensional diffeomorphisms
and $U(1)$ gaume symmetry $(\xi^{\hmu},c^*)$ are trivially mapped as can be
seen by inspection of equations (\ref{gf1})-(\ref{gf4}). As we already know
from the bosonic analysis, $V_\rho$ becomes the new T-dual scalar, as can be
seen from its diffeomorphism transformation (\ref{gf3}).

\subsection{Kappa symmetry}
\indent

In Appendix \ref{kappaproof}, it is proved that
\begin{equation}
{\delta_\kappa {\7\theta}^\ua |_\CA  ~~~
\limit{\longrightarrow}{\CT_{\parallel}}~~~ 
\delta_{\kappa'}{\7{\theta'}}^\ua },
\label{k1}
\end{equation}
where IIA and IIB kappa parameter functions are related by the same form of 
equations as $\T$'s in \bref{tch2}
\bea
\7\K_1&=&a_2~\7\K'_2,~~~~~~
\7\K_2~=~-a_1~\G_\uz~\7\K'_1.
\label{tch2k}
\eea
It shows that kappa symmetry transformations of the fermionic
sector of the theory are correctly mapped under T-duality. 
Once (\ref{k1}) is known, it is straightforward to extend the proof
for $\delta_\K \7x^{\8{\uaa}}$, using the identities (\ref{conditions}).
The first non-trivial check is the gauge symmetry analysis of
the T-duality mapping $V_\rho |_\CA=\h z'$ in \bref{cons},
\bea 
\delta_\kappa V_\rho|_\CA = -
\delta_\kappa\5{\7\theta} \Gamma_{11}\Gamma_\uz {\7\theta}~\Lam +
\delta_\kappa x^{\8m} b_{\8m z} \label{gf31} ~~~
\limit{\longrightarrow}{\CT_{\parallel}}~~~
\eta \delta_\K z' ~=~
\Lam\delta_{\kappa'}{\5{\7{\theta'}}}
\G^{\uz}{\7{\theta'}} + 
{\Lam}\delta_{\kappa'}x'^{\8m}{e'_{\8m}}^{\uz}.
\eea
It can be seen that $\delta_\K V_\rho |_\CA$ turns out
to be the kappa symmetry transformation for the T-dual scalar 
$\delta_{\kappa'}z'$, thus allowing us to write the kappa symmetry
transformations of the bosonic scalar sector in the T-dual description
in a fully ten dimensional covariant way 
\bea
\delta_{\kappa'} \7x'^{\uaa} = -\delta_{\kappa'}{\5{\7{\theta'}}}\G^{\uaa}
{\7{\theta'}},~~~~~~~{\rm for} ~~~~~~\uaa=(\8\uaa,\uz).
\eea

We are left with kappa transformations of the $V_{\hmu}$ components.
There is an additional contribution to the kappa transformation from
$\xi^\rho$ in \bref{comp}. 
Using the identities (\ref{conditions}), it can be shown that
\begin{equation} 
\delta^*_\kappa V_{\hmu}\equiv \left(\delta_\kappa V_{\hmu} + 
\delta_\kappa z \6_{\hmu} V_\rho \right)
\limit{\longrightarrow}{\CT_{\parallel}}
\delta_{\kappa'} V'_{\hmu}, 
\label{k3}
\end{equation}
where 
\bea
\delta_{\K'} V'_\hmu =-\delta_{\kappa'} \5{\7\theta'}\tau_{3}
\Gamma_\uaa\7\T'\left(\6_\hmu \7x^{'\uaa} - {1\over 2}\5{\7\theta'}
\Gamma^\uaa \6_\hmu {\7\theta'}\right) 
 +{1\over 2}\delta_{\kappa'}\5{\7\theta'}\Gamma^\uaa{\7\theta'}
\5{\7\theta'}\tau_{3}\Gamma_\uaa\6_\hmu {\7\theta'}
+\D_{\kappa'} {x'}^m\6_\hmu {x'}^nb'_{mn}  
\nn\\
\label{gv1}
\eea
which finishes the proof of our claim.

\subsection{Supersymmetry}
\indent

It is natural to apply the T-duality transformation properties
of the fermionic scalar fields 
(\ref{tch2}) for the supersymmetry parameters
\bea
\7\ep_1&=&a_2~\7\ep'_2,~~~~~~
\7\ep_2~=~-a_1~\G_\uz~\7\ep'_1\,.
\label{tch2a}
\eea
In this way 
\begin{equation} 
\delta_\ep {\7\theta}^\ua |_\CA~~
\limit{\longrightarrow}{\CT_{\parallel}}~~
\delta_{\ep'}{\7{\theta'}}^\ua 
\label{k1a}
\end{equation}
and the corresponding behaviour for $\delta_\ep \7x^{\8\uaa}$ 
follows immediately. We are thus left with the supersymmetry
transformations of the gauge field. To begin with,
\bea
\delta_\ep V_\rho = \5{\7\epsilon}\Gamma_{11}\Gamma_\uz {\7\theta}~\Lam
+ \D_\ep x^{\8m} b_{\8m z} \label{gfg11}~~~
\limit{\longrightarrow}{\CT_{\parallel}}~~~
 \h \delta_\ep z'=
-\Lam(\5{\7\ep'}\G^{\uz}\7\theta' -
\5{\7\ep'}\G^{\8\uaa}\7\theta' \,e'^{\8m}_{\8\uaa} \,e'^{\uz}_{\8m})
\eea
from which the ten dimensional character of $V_\rho$ can be emphasized
again and
\bea
\delta_{\ep'} \7x'^{\uaa} = \delta_{\ep'}{\5{\7{\theta'}}}\G^{\uaa}
{\7{\theta'}}~~~~~~~{\rm for} ~~~~~~\uaa=(\8\uaa,\uz).
\eea

Concerning to $V_\hmu$ the susy transformation modified by $\xi^\rho$
is mapped to the IIB one as in the kappa symmetry case,
\begin{equation} 
\delta^*_\ep V_{\hmu}\equiv \left(\delta_\ep V_{\hmu} + \delta_\ep z 
\6_{\hmu} V_\rho \right)~~
\limit{\longrightarrow}{\CT_{\parallel}}~~
\delta_{\ep'} V'_{\hmu} \,.
\label{s3}
\end{equation}

\subsection{Poincare Bosonic global symmetries}
\indent

Let us concentrate on the manifest $ISO(1,8)$ symmetry group. From the
transformation of $\7x^{\8\uaa}$ in \bref{glgfx} and $V_\rho$ in \bref{gfg1} 
it follows
\bea
\Delta {{\7x}}^{'\8\uaa}~=~{\rm a}^{\8\uaa}+\w^{\8\uaa}_{~{\8\ubb}}x^{\8\ubb}
,~~~~~~~~
\Delta {{\7x}}^{'\uz}&=&A\Lam',
\eea
which allows us to interpret it as the corresponding $ISO(1,8)$
infinitesimal transformations 
and a ${{\7x}}^{'\uz}$ coordinate translation in the T-dual target space, 
whenever we take the constant
$A$ as $A=\frac{{\rm a}'^{\uz}}{\Lam'}$, without loss of generality. It is
worthwhile emphasizing, as in \cite{joan}, that the origin of the
translational symmetry is the original $U(1)$ gauge symmetry.
Furthermore
\begin{equation}
\delta^*_{\rm a} V_{\hmu} + \delta^*_{\omega} V_{\hmu} 
\longrightarrow 
\delta_{{\rm a}'} V'_{\hmu} + \delta_{\omega'} V'_{\hmu} +
\6_{\hmu} c_{(1)}
\end{equation}
does describe the $ISO(1,8)$ transformations in the T-dual theory
up to a $U(1)$ transformation, which can be absorbed in a redefinition
of the T-dual $U(1)$ gauge parameter $c^*$ without loss of generality.

The latter analysis shows the existence of the $ISO(1,8)$ symmetry
group, but we know it should be enhanced to the full $ISO(1,9)$.
In fact, there is no theorem guaranteeing the equality of the full
symmetry group of the T-dual effective action with the symmetry group
of the partially gauge fixed truncated action. What is indeed true
is that the latter group is a subgroup of the former. In other words,
$H^{-1,p+1}(s|d)\subseteq H^{-1,p}(s_{{\cal A}}|d_{{\cal A}})$,
$H^{-1,n}(s|d)$ being the cohomological group at ghost number minus one
characterizing the set of non-trivial global symmetries of any n-dimensional
classical field theory \cite{brandt}. There are examples of such an
enhancement in the literature. For instance, the D-string effective action
is known to have an infinite set of non-trivial global symmetries
\cite{ours}, while such structure is not known to exist for the D2-brane
effective action, even though they are T-dual to each other. In the
present case, it is certainly true that the T-dual theory is invariant
under the following set of rotations, 
\bea
\Delta \7x'^{\8{\uaa}} &=& \omega'^{\8{\uaa}\uz}\7x'^{{\uz}}, ~~~~~~~
\Delta \7x'^\uz ~=~  \omega'^{\uz}_{~~\8{\uaa}}\7x'^{\8{\uaa}} \\
\Delta \7\theta'^\ua &=& \frac{1}{2}\omega'^{\uz\8{\ubb}}
\left(\G_{\uz\8{\ubb}}\7\theta'\right)^\ua \\
\Delta V'_{\hmu} &=& \Delta x'^m \6_{\hmu}x'^n {b'_{mn}}
\eea
as is clear from the fact that
the T-dual IIB action has manifest $ISO(1,9)$ invariance.  

%%%%%%%%%%%%%%%%%%%%%%%%%%%%%%%%%%%%%%%%%%%%%%%%%%%%%%%%%%%%%%%%%%%%%%%%%%%%%5
\section{Supersymmetric world volume solitons}
\indent

The goal of the present section is to prove that any bosonic 
supersymmetric world volume soliton of a IIA D$p$-brane in constant
$G_{mn}$, $b_{mn}$ backgrounds is mapped under T-duality into the 
corresponding bosonic supersymmetric world volume configuration for 
the T-dual theory. This will be shown in two different, but complementary,
ways. First of all, the T-duality behaviour of the kappa symmetry
preserving condition will be analyzed, and after that, the same analysis
will be carried in the hamiltonian formalism describing D-branes
\cite{paul}, paying special attention into the hamiltonian constraint,
giving rise to the energy density of such BPS configurations.

It is known that any bosonic supersymmetric world volume configuration
must satisfy \cite{ertg}
\begin{equation}
\Gamma_\K \ep = \ep,
\label{spc}
\end{equation}
where $\Gamma_\K$ is the matrix appearing in the kappa symmetry transformations
$(\delta_\kappa \7\theta^\A)$ while $\ep$ is the Killing spinor of the 
corresponding background geometry, in our case a constant spinor. It will be
useful to determine $\G_\kappa$ explicitly, in terms of the $\gamma^{(p)}$
matrix appearing in our previous discussions of kappa symmetry. Due to the fact
that $\Ttb=\Tt^tC$, where $C$ is the antisymmetric charge conjugation
matrix, it is straightforward to derive the type IIA relation
\bea
\G_\kappa &=&C^{-1}\gam^{(p)t}C~=~
\frac{1}{\sqrt{-det(\CG+\CF)}}
[\sum_{\l=0}\frac{{\slPi^{2\l+1}}(\ggt)^{\l+1}}{(2\l+1)!}~e^\CF]_{p+1}
\eea
whereas for type IIB  \Dpm-brane, it reads as
\begin{eqnarray}
\G'_{\kappa'}&=&\frac{1}{\sqrt{-det(\CG'+\CF')}}
[\sum_{\l=0}\frac{(\slPi)^{'2\l}}{(2\l)!}(\tau_3)^{\l+1}\tau_1~e^{\CF'}
]_{p}.
\eea 

\subsection{BPS equations and susy projectors}
\indent

It will be proved that the kappa symmetry preserving condition (\ref{spc}),
when projected into the subspace $\CA$ and applying a T-duality transformation
on the background $({e_m}^{\underline{a}},b_{mn})$, dynamical fields
$(\phi^i)$ and Killing spinor $(\epsilon)$, is correctly mapped into the
corresponding kappa symmetry preserving condition in the T-dual theory
$\Gamma_{\kappa'}\ep'=\ep'$.
Consider equation (\ref{spc}) and split it into its different chiral
components
\bea
\ep_1&=&\frac{1}{\sqrt{-det(\CG+\CF)}}
[\sum_{\l=0}\frac{(-1)^{\l+1}{\slPi^{2\l+1}}}{(2\l+1)!}~e^\CF]_{p+1}\ep_2,
\nn\\
\ep_2&=&\frac{1}{\sqrt{-det(\CG+\CF)}}
[\sum_{\l=0}\frac{{\slPi^{2\l+1}}}{(2\l+1)!}~e^\CF]_{p+1}\ep_1\, .
\label{gsusypro}
\eea
Using the T-duality relations 
\bea
\ep_1~=~a_2 \ep'_2,~~~~~~\ep_2~=~-~a_1 \G_\uz \ep'_1\,,
\label{aepsrel3}
\eea
and the T-duality transformation properties of the matrix $\Gamma_{\kappa}$, 
equations (\ref{gsusypro}) can be rewritten as
\bea
a_2 \ep'_2
&=&\frac{a_1~}{\sqrt{-det(\CG'+\CF')}}
[\sum_{\l=0}\{(-1)^{\l}\frac{(\slPi')^{2\l}}{(2\l)!}\}
e^{\CF'}]_{p }\ep_1',
\\
-a_1 \ep'_1
&=&\frac{a_2}{\sqrt{-det(\CG'+\CF')}}
[\sum_{\l=0}\{\frac{(\slPi')^{2\l}}{(2\l)!}\}
e^{\CF'}]_{p }~\ep_2'.
\eea
which are combined to give the final result 
\bea
\ep'&=&\frac{1}{\sqrt{-det(\CG'+\CF')}}
[\sum_{\l=0}\{\frac{(\slPi')^{2\l}}{(2\l)!}(\tau_3)^{\l+1}\}\tau_1~
e^{\CF'}]_{p }~\ep'~=~\G'_{\kappa'}~\ep'.
\nn\\
\eea
It is important to stress that the latter proof applies to any
configuration solving the kappa symmetry preserving condition
in the subspace $\CA$. Since any solution to such a condition involves 
a set of BPS equations and a set of supersymmetry projection conditions, we 
conclude that both sets of equations are mapped to the corresponding BPS 
equations and supersymmetry projection conditions under T-duality,
thus generating a supersymmetric world volume soliton for the T-dual 
theory. This is nothing but the same phenomena observed in supergravity
theories describing the low energy dynamics of the massless
closed string spectrum. There, T-duality is a generating solution
transformation. The above proof, which will be examined
in particular examples in next subsections, ensures the same
generating character for the low energy dynamics of the massless
open string spectrum.

\subsection{Hamiltonian analysis}
\indent

Given any world volume brane theory, and for any bosonic supersymmetric 
configuration solving (\ref{spc}), one can always use its phase space 
formulation to compute its energy density \cite{quim}. When the world volume 
theory is defined on a SuperPoincar\'e background, it gives us a field theory 
realization of the SuperPoincar\'e algebra. Since BPS states are known to 
saturate the BPS bound, it must always be possible to write the energy
density as a sum of squares,
\begin{eqnarray}
{\cal E}^2 &=& {\cal E}^2_0+ \CZ^2 + \sum_i \left(t^i f_i(\phi^j)\right)^2 \,,
\label{nonthreshold} 
\eea
for non-threshold  BPS states and
\bea
{\cal E}^2 &=& ({\cal E}_0+ \CZ)^2 + \sum_i \left(t^i f_i(\phi^j)\right)^2 \,,
\label{threshold}
\end{eqnarray}
for threshold  BPS states.
Here ${\cal E}_0$ stands for the energy density of the vacuum configuration
and $f_i(\phi^j)=0$ stand for the set of BPS equations derived from
(\ref{spc}). 
They allow us to define a natural lower bound on the energy or {\em BPS 
bound} respectively,
\begin{eqnarray}
{\cal E} &\geq& \sqrt{{\cal E}^2_0+ \CZ^2}, \label{nonthresholdbound} \\
{\cal E} &\geq& {\cal E}_0 + |\CZ| \label{thresholdbound}
\end{eqnarray}
being saturated precisely when $f_i(\phi^j)=0$ are satisfied, thus justifying
their qualification as {\em BPS equations}\footnote{We have assumed the
existence of a single $\CZ$ charge in the above derivation, but the extension
to more general configurations is straightforward and completely analogous
to the BPS bounds derived from a pure algebraic approach.}.

If our previous analysis was correct, it should be possible to prove
that the hamiltonian constraint of the original theory is mapped to the
T-dual hamiltonian constraint in the T-dual theory. To prove this we will study
the phase space description of D-branes in 
constant $G_{mn}$ and $b_{mn}$ backgrounds by setting $\theta=0$. 
The phase space formulation is given by
\cite{paul}
\begin{eqnarray}
\CL & = & P_m\dot{x}^m + E^a\dot{V}_a + V_t\partial_a E^a -
s^a\left(\7P_{\uaa}\Pi^{\uaa}_a + E^{b}\CF_{ab}\right) \nonumber \\
& & -{1\over 2}v\left[\7P^2 + E^{a}E^{b} g_{ab} + T_p^2\,
det\,(\CG_{ab} + \CF_{ab})\right],
\label{ham}
\end{eqnarray}
where $\CG_{ab}$ stands for the world space induced metric.
$P_m$ and $E^a$ are conjugate momenta of $x^m$ and $V_a$ respectively and 
%$\Pi^{\uaa}_a=\6_a x^m e_m~^{\uaa}$ and
$P_m = e_m~^{\uaa} \7P_{\uaa} - E^a\6_a x^n b_{mn}$. When computed
in the subspace $\CA$,
\bea
\Pi^{{\underline{z}}}_{\8a} &=& \6_{\8a}x^{\8m} e_{\8m}~^{{\underline{z}}} 
\, ,~~~~~~~~ \, \Pi^{{\underline{z}}}_\rho = \lambda \\
\Pi^{\8\uaa}_{\8a} &=& \6_{\8a}x^{\8m} e_{\8m}~^{\8\uaa} \, , ~~~~~~~~\,
\Pi^{\8\uaa}_\rho =0 \\
P_{\8m} & = & e_{\8m}~^{\8\uaa}\7P_{\8\uaa} + e_{\8m}~^{{\underline{z}}}
\7P_{{\underline{z}}} - E^{a}\6_{a}x^{\8n} b_{\8m\8n} - E^\rho 
b_{\8mz} \\
P_z &=& \lambda \7P_{{\underline{z}}} + E^{\8a}\6_{\8a}x^{\8m} b_{\8mz}.
\eea
Note that $a,b$ stand for world space indices, while the underlined ones 
stand for background tangent space indices. 

It is straightforward to derive the T-duality properties of these
objects from the rules that we have already derived in the lagrangian
formulation :
\bea
  \Pi^{\8\uaa}_{\8a} &\longrightarrow& \Pi'^{\8\uaa}_{\8a}   \\
  \Pi^{{\underline{z}}}_{\8a} &\longrightarrow& \6_{\8a}x^{\8m} 
{b'_{\8mz}\over \lambda'}   \\
  \CF_{\8a\rho} &\longrightarrow& {\eta\over \lambda'} 
\Pi'^{{\underline{z}}}_{\8a}   \\
  \CF_{\8a\8b} &\longrightarrow& \CF'_{\8a\8b} + \6_{\8a}x^{\8m} 
{b'_{\8mz}\over 
\lambda'} \Pi'^{{\underline{z}}}_{\8b} - \Pi'^{{\underline{z}}}_{\8a}
\6_{\8b}x^{\8m} {b'_{\8mz}\over \lambda'} \\
det\,(\CG_{ab}+\CF_{ab}) & = & \lambda^2\,
det\,(\CG^\prime_{\8a\8b}+\CF^\prime_{\8a\8b}).
\eea
Note that $\CL'=\int d\rho \CL$. Let us be more specific and rewrite
$\CL=T_p \8\CL$, $\CL'=T_{p-1}\8\CL'$. It is clear that under T-duality
$\8\CL=\Lam \8\CL'$. From these considerations, we can derive the
following relation between momenta
\begin{eqnarray}
P_{\8m}&=&T_p \frac{\6{\8\CL}}{\6\dot{x}^{\8m}} = \frac{\lambda T_p}{T_{p-1}}
\Gamma_{\8m}~^{\8n}P'_{\8n} \quad
, \quad P'_{\8n} = T_{p-1} \frac{\6\8\CL'}{\6\dot{x}'^{\8n}} \nonumber \\
E^\rho & = &  {\lambda T_p \over \eta T_{p-1}}P^\prime_{\7z} \nonumber \\
E^{\8a} & = & {\lambda T_p \over T_{p-1}} E^{'\8a} 
\label{au}
\end{eqnarray}
from which we can derive that $\7P_{\8\uaa} = {\lambda T_p \over
T_{p-1}}\7P'_{\8\uaa}$.

The process of partial gauge fixing $(z=\rho)$ in the lagrangian formulation
corresponds, in the phase space formulation, to solve the equation
\begin{equation} 
\frac{\delta\CL}{\delta s^\rho}=0~~~ \Longrightarrow~~~  
\7P_{{\underline{z}}} = {E^{\8b}\CF_{\8b\rho} \over \lambda}.
\label{pz}
\end{equation}
Using the above information, one can show that the remaining diffeomorphism
constraints $(\delta\CL/\delta s^{\8a}=0)$ and the hamiltonian constraint
$(\delta\CL/\delta v=0)$
are mapped to the corresponding T-dual constraints, by defining 
$v' = {\lambda T_p \over T_{p-1}} v$,
\begin{eqnarray}
\frac{\delta\CL}{\delta s^{\8a}} \longrightarrow \frac{\delta\CL'}{\delta
s'^{\8a}} \nonumber \\
\frac{\delta\CL}{\delta v} \longrightarrow \frac{\delta\CL'}{\delta v'}\,,
\end{eqnarray}
thus indeed proving our initial claim.

\subsection{Examples}
\indent

The aim of this subsection is to show, explicitly, how the BPS equations
and supersymmetry projection conditions characterizing world volume
solitons are mapped into the corresponding ones under T-duality. 
We will, first of all, concentrate on orthogonal BIon solutions that are 
common to all D$p$-branes and on dyons in a D3-brane propagating in 
SuperPoincar\'e background $(b_{mn}=0)$. After that, we consider
the more subtle effect of T-duality on tilted BIons $(b_{mn}\neq 0)$.
In the following we will be using the explicit parametrisation
$\eta=a_2=-a_1=1$ and $\Gamma_m~^n=\delta_m~^n$ which can always
be done.

\subsubsection{BIons and dyons}
\indent

Let us start our discussion with {\em BIons}. That is, we will look for 
classical solutions to the D$p$-brane equations of motion propagating in 
Minkowski space, corresponding to a fundamental string ending on the brane. 
The latter configuration is known to be described by the ansatz \cite{callan}
\begin{equation}
x^\mu = \sigma^\mu \quad , \quad x^{p+1}=y(\sigma^a) \quad ,
\quad V_0 = V_0(\sigma^a) \, ,
\label{ans1}
\end{equation}
$\mu=0,\ldots ,p$, $a=1,\ldots ,p$, the rest of bosonic scalar fields being 
constant and the magnetic components of the gauge field have %are set to be
a pure gauge configuration, corresponding to the array of branes
\begin{equation}
\begin{array}{ccccccccccl}
Dp: &1&2&.&.&p&\_&\_&\_&\_ &\quad \mbox{probe}    \nonumber \\
F1: &\_&\_&\_&\_&\_&y&\_&\_&\_ &\quad \mbox{soliton.} 
\end{array}
\end{equation}
The solution to the kappa symmetry preserving condition when (\ref{ans1}) is 
satisfied, is given by
\begin{eqnarray}
\CP_1\ep & = & \ep \label{a1} \\
\CP_2\ep & = & \ep \label{a2} \\
F_{0a} & = & \partial_a y \, ,\label{a3}
\end{eqnarray}
where $\ep$ is a constant Killing spinor. The first two conditions
(\ref{a1}-\ref{a2}) correspond to the supersymmetry projection conditions
telling us that we are describing a D$p$-brane and a fundamental string, 
$\CP_2 = \G_{\underline{0}\underline{y}}\G_{11}$ in type IIA and
$\CP_2 = \G_{\underline{0}\underline{y}}\tau_3$ in type IIB, while equation 
(\ref{a3}) is the usual BPS equation, which
by using the Gauss' law $(\6_a E^a = \6_a \delta^{ab}F_{0b}=0)$ determines 
the harmonic character of the excited transverse scalar 
$(\delta^{ab}\6_a\6_b y =0)$. 

Let us study the effect of T-duality along the world volume 
direction $\rho=p$. If, as suggested by our analysis, we apply the partial 
gauge fixing plus functional truncation on (\ref{a3}), the only non-trivial
equation that we get is the corresponding BPS equation in the T-dual
description
\begin{equation}
F'_{0\uaa}=\partial_{\uaa} y'\, .
\label{bb}
\end{equation}

Concerning supersymmetry projections, take eq. (\ref{a1}) for a D4-brane.
In that case, $\CP_1 = \G_{\underline{0}\underline{1}\underline{2}
\underline{3}\underline{4}}\G_{11}$ and eq. (\ref{a1}) is equivalent to 
\bea
\G_{\underline{0}\underline{1}\underline{2}
\underline{3}\underline{4}}\ep_2 &=& -\ep_1 \nn \\
\G_{\underline{0}\underline{1}\underline{2}
\underline{3}\underline{4}}\ep_1 &=& \ep_2
\label{p1}
\eea
which when written in terms of the T-dual Killing spinors $\ep'$ look as
\begin{eqnarray}
\G_{\underline{0}\underline{1}\underline{2}\underline{3}}\ep'_1 &=& 
-\ep'_2 \nn \\
\G_{\underline{0}\underline{1}\underline{2}\underline{3}}\ep'_2 &=& 
\ep'_1
\label{p2}
\end{eqnarray}
which is consistent with the projection 
$\G_{\underline{0}\underline{1}\underline{2}\underline{3}}i\tau_2\ep^\prime=
\ep^\prime$ satisfied by a D3-brane. Concerning the soliton 
projection (\ref{a2}), it can be splitted into
\begin{eqnarray}
\G_{\underline{0}\underline{y}}\ep_1 & = & -\ep_1 \\
\G_{\underline{0}\underline{y}}\ep_2 & = & \ep_2 \, ,
\end{eqnarray}
which are equivalent to
\begin{eqnarray}
\G_{\underline{0}\underline{y}}\ep^\prime_2 & = & \ep^\prime_2 \\
\G_{\underline{0}\underline{y}}\ep^\prime_1 & = & -\ep^\prime_1 \, ,
\label{a4}
\end{eqnarray}
respectively. Equations (\ref{a4}) can be joined into 
$\G_{\underline{0}\underline{y}}\tau_3\ep^\prime =-\ep^\prime$ \footnote{The
minus sign is related with the freedom of choosing as a BPS equation
$F'_{0\underline{a}}=-\6_{\underline{a}}y'$, instead of (\ref{bb}).}, which is 
the soliton projection for a fundamental string in type IIB. Analogous 
discussion applies for other values of $p$.

To sum up, we have indeed shown that BPS equations and susy projections
are mapped, under T-duality, to the corresponding BPS equations and
susy projections describing the T-dual configuration
\begin{equation}
\begin{array}{ccccccccccl}
D(p-1): &1&2&.&p-1&\_&\_&\_&\_&\_ &\quad \mbox{probe}    \nonumber \\
F1: &\_&\_&\_&\_&\_&y&\_&\_&\_ &\quad \mbox{soliton.} 
\end{array}
\end{equation}

We would like to comment on the explicit solution of the harmonic
equation $\delta^{ab}\6_a \6_b y=0$. Of course, we could just restrict
ourselves to a particular solution of this equation 
%, not involving an explicit 
independent on the world volume coordinate $\rho$ along
which we are T-dualizing, to be consistent with the functional
truncation we were discussing in previous sections. Another possibility is to
consider a superposition of BIons of the same mass and charge, located 
periodically along the {\boldmath$\8\rho$} axis with period $a=2\pi R$
\cite{perry},
\[y = k_p \sum_{n\in{\bZ}} {1\over |\sigma - na\8\rho|^{p-2}}
\quad p \geq 3 \]
\[y =  k_2 \sum_{n \in \bZ} \log |\sigma - na\8\rho| \quad p=2 \, .\]

In the limit $R \longrightarrow 0$, which is the one we have been
studying along the whole paper, the discrete sum is replaced by an integral,
\[ \sum_{n \in \bZ} {k_p\over |\sigma - na\8\rho|^{p-2}} \longrightarrow
\int^\infty_{-\infty} {k_p d\rho \over (\8\sigma^2 + \rho^2)^{(p-2)/2}}
= {\7k_p \over \8\sigma^{p-3}} \quad p\geq 4 \]
\[ \sum_{n \in \bZ} {k_3\over |\sigma - na\8\rho|} \longrightarrow
\int^\infty_{-\infty} {k_3 d\rho \over (\8\sigma^2 + \rho^2)^{1/2}}=
\7k_2 \log |\8\sigma| \quad p=3 \]
\[ \sum_{n \in \bZ} k_2\log |\sigma - na\8\rho| \longrightarrow
\int^\infty_{-\infty} k_2d\rho \log |\sqrt{\sigma^2 + \rho^2}| = \7k_1 \sigma_1
\quad p=2 \]
which is effectively equal to ignoring all the heavy modes along the 
{\boldmath$\8\rho$} direction, giving the correct functional behaviours
in the T-dual theory.

Let us now describe the effect of T-duality on {\em dyons}. We will look for 
classical solutions to the D3-brane equations of motion propagating in 
Minkowski space, corresponding to a $(p,q)$ string ending on the brane. 
The latter configuration is known to be described by the ansatz \cite{quim}
\begin{equation}
x^\mu = \sigma^\mu \quad , \quad x^{p+1}=y(\sigma^a) \quad ,
\quad V_0 = V_0(\sigma^a) \, , \, V_b = V_b(\sigma^a) \, ,
\label{ans2}
\end{equation}
where $\mu=0,\ldots ,3$, $a,b=1,\ldots ,3$, and 
the rest of bosonic fields are being 
constant, corresponding to the array of branes
\begin{equation}
\begin{array}{ccccccccccl}
D3: &1&2&3&\_&\_&\_&\_&\_&\_ &\quad \mbox{probe}    \nonumber \\
F1: &\_&\_&\_&4&\_&\_&\_&\_&\_ &\quad \mbox{soliton}     \nonumber \\
D1: &\_&\_&\_&4&\_&\_&\_&\_&\_&\quad \mbox{soliton.}
\end{array}
\label{IIB}
\end{equation}
The solution to the corresponding kappa symmetry preserving condition is 
\begin{eqnarray}
\G_{\underline{0}\underline{1}\underline{2}\underline{3}}\,i\tau_2\ep & = & 
\ep \label{b1} \\
\G_{\underline{0}\underline{y}}\left(\cos\alpha\,\tau_3 + \sin\alpha\,
\tau_1\right)\ep & = & \ep \label{b2} \\
F_{0a} & = & \cos\alpha\,\partial_a y \label{b3} \\
{1\over 2}\ep^{abc}F_{bc} & = & \sin\alpha\,\delta^{ab}\partial_b y.
\label{b4}
\end{eqnarray}
Equations (\ref{b1})-(\ref{b2}) are the supersymmetry projection
conditions for this configuration. The first one describes a D3-brane along
directions $123$, as expected, while the second one describes a $(p,q)-string$
along the transverse direction $y$. Equations (\ref{b3})-(\ref{b4})
are the BPS equations for this configuration \cite{quim}.

The longitudinal T-dual configuration is known to be
\begin{equation}
\begin{array}{ccccccccccl}
D2: &1&2&\_&\_&\_&\_&\_&\_&\_ &\quad \mbox{probe}    \nonumber \\
F1: &\_&\_&\_&4&\_&\_&\_&\_&\_ &\quad \mbox{soliton}     \nonumber \\
D2: &\_&\_&3&4&\_&\_&\_&\_&\_ &\quad \mbox{soliton.}
\end{array}
\label{IIA}
\end{equation}
Proceeding as before, the truncated BPS equations one gets are 
\begin{eqnarray}
F'_{0\8a} & = & \cos\alpha\,\partial_{\8a} y' \label{b5} \\
\ep^{\8a\8b}\partial_{\8b}z' & = & \sin\alpha\,\delta^{\8a\8b}\partial_{\8b}y'
\label{b6}
\end{eqnarray}
while the supersymmetry projections become 
\bea
\G_{\underline{0}\underline{1}\underline{2}}\ep' &=& \ep' \label{b7} \\
\left(-\G_{\underline{0}\underline{y}}\G_{11}\cos\alpha+\G_{\underline{0}
\underline{y}\underline{z}}\sin\alpha\right)\ep'&=& \ep'\, .
\label{b8}
\eea
Equations (\ref{b5}-\ref{b8}) describe a threshold bound state of a 
D2 brane and a fundamental IIA string realized on the world volume of the 
first D2-brane. Note that studying the particular limit, $\alpha=0$ we
recover the BIon discussion, while for $\alpha=\pi/2$, equation
(\ref{b6}) is equivalent to the Cauchy-Riemann equations (when written
in terms of complex world volume coordinates and the complex function
$U=y+iz$) describing a $D2\perp D2(0)$, which is the direct dimensional
reduction of the $M2\perp M2(0)$ configuration \cite{quim,gary1}.

\subsubsection{World volume solitons in constant b fields}
\indent

We will concentrate on world volume solitons on a D3-brane proving in
Minkowski space
and a constant arbitrary $b_{mn}$ field. The kappa symmetry preserving
condition looks as
\bea
\sqrt{-det\,(\CG + \CF)} \ep &=& {1\over 4!}\ep^{\mu_1\ldots\mu_4}
\left(\gam_{\mu_1\ldots\mu_4}\,i\tau_2 + 6\,\CF_{\mu_1\mu_2}\,
\gam_{\mu_3\mu_4}\,\tau_1 + 3 \CF_{\mu_1\mu_2}\CF_{\mu_3\mu_4}\,
i\tau_2\right)\epsilon . \nonumber \\
\label{kp1}
\eea
We will describe two different configurations solving eq.(\ref{kp1}). 
First of all, we will generalize the BPS equations (\ref{b3}-\ref{b4})
describing {\em dyons} in the absence of a $b_{mn}$ field. Using the same
ansatz as in (\ref{ans2}), the solution to (\ref{kp1}) involves
the same supersymmetry projectors (\ref{b1}) and (\ref{b2}), while
the BPS equations are given by
\bea
\CF_{0a}&=&\cos\A\,\6_a y \label{c3} \\
{\cal B}^a &=& {1\over 2}\ep^{abc}\CF_{bc} = \sin\A~\delta^{ab}\6_b y
\label{c4}
\eea
which are the straightforward generalization of the usual dyonic BPS equations
in the presence of a $b_{mn}$ field, this being the reason of the appearance
of the gauge invariant tensor $\CF$. 

If we T-dualize along the direction $3$, the BPS equations that we obtain 
are
\bea
\6_0 z' &=& -G'_{03} \label{d1} \\
\CF'_{0\uaa} &=& \cos\A\,\6_{\uaa} y' \label{d2} \\
{1\over 2} \ep^{\8a\8b}\CF_{\8a\8b} &=& 0 \label{d3} \\
\ep^{\8a\8b}\left(\6_{\8b}z' + G'_{\8b 3} +
\6_{\8b}y' G'_{\8b y}\right) &=& \sin\A \6_{\8a} y' \,,
\label{d4}
\eea
where $\8a,\8b=1,2$. 
The most remarkable feature of this T-dual configuration
is being non-static, see equation (\ref{d1}). Let us discuss in more detail
equations (\ref{c3}) and (\ref{c4}) when the background is such that only
the electric components of the $b_{mn}$ field along the world volume are 
non-vanishing $(b_{0a}\neq 0)$, and $\A=0$, that is, we will be concerned
with {\em BIon} type solutions. In this case, eqs. (\ref{c3}) and (\ref{c4})
besides the Gauss' law are easily integrated to give the solution
\bea
y(\sigma^b)&=& y_{h}(\sigma^b) + d_a\sigma^a \label{s1} \\
V_0(\sigma^b) &=& -y_h(\sigma^b) + c_a\sigma^a \label{s2} \\
d_a + c_a &=& -b_{0a} \label{s31} \, ,
\eea
where $y_h(\sigma^b)$ denotes the harmonic part of the solution 
whereas $d_a,c_a$ is some set of constants constrained by (\ref{s31}). 
Notice that $d_a$ are physical parameters, due to the gauge invariant 
character of the excited scalar, determining the tilting of the BIon 
\cite{hashi}. In other words, due
to the non-orthogonal character of the BIon, when we study T-duality
along the $\rho$ direction, this is seen as a T-duality at angle from
the BIon perspective. As a result of that, one should expect the
T-dual configuration to be one with a tilted BIon ending on a D2-brane
boosted in the direction of dualization, which is what we get
from inspection of eq.(\ref{d1}). 
To sum up, constant electric field $(\CF_{0a})$ 
boosts the configuration in the direction along we T-dualize
, just as constant flux of 
magnetic field on the D-brane $(\CF_{ab})$ is seen as D-branes at angles 
in the T-dual picture \cite{Pol}. 

Due to the generating character of the T-duality transformation, 
we would like to understand how 
the original static configuration give rise to
the non-static solution in the T-dual theory. 
%we would like to understand how it is 
%that the original configuration was static,
%while the T-dual one is not. 
First of all, due to the functional truncation
defining the ${\cal A}$ subspace, we will choose $d_3=0$, thus avoiding
the linear dependence in $\sigma^3$ on the gauge invariant quantity
$y(\sigma^b)$. By (\ref{s31}), $c_3=-b_{03}$, or equivalently,
\begin{equation}
V_0(\sigma^b) = -y_h(\sigma^b) + c_{\8a}\sigma^{\8a} -b_{03}\sigma^3\,.
\label{mn}
\end{equation}
The latter has also a linear dependence in $\sigma^3$, but this is certainly
gauge dependent, since we can find a gauge parameter $c=b_{03}\tau\sigma^3$
transforming the gauge field configuration (\ref{mn}) into
\begin{eqnarray}
V_0(\sigma^b) &=& -y_h(\sigma^b) + c_{\8a}\sigma^{\8a} \nn \\
V_3(\tau) &=& b_{03}\tau\,,
\end{eqnarray}
which is explicitly time dependent. The latter is the most natural
higher dimensional solution giving rise to the non-static
T-dual configuration \footnote{JS would like to thank David Mateos
for discussions related to this point.}.

As a second example, we will consider a non-threshold bound state of a
D-string inside the D3-brane together with some BIon, which is generically
tilted, due to the non-vanishing of the $b$ field. Using the same
ansatz as in (\ref{ans2}), the  solution to the kappa symmetry preserving 
condition is given by
\bea
\left(\cos\A\Gamma_{\underline{0}\underline{1}\underline{2}
\underline{3}}\,i\tau_2 + \sin\A\Gamma_{\underline{0}\underline{1}}\,\tau_1
\right) \ep &=& \ep \label{tbs} \\
\Gamma_{\underline{0}\underline{4}}\,\tau_3 \ep &=& \ep \label{spike} \\
\CF_{23} = \CF &=& \tan \A \label{tbs1} \\
\CF_{0\uaa} &=& -\6_{\uaa} y \quad \uaa = 2,3 \label{spike1} \\
\CF_{01} = \6_1 y &=& 0. \label{spike2}
\eea
Equations (\ref{tbs}), (\ref{tbs1}) are a straightforward generalization
of the conditions satisfied by any non-threshold bound state involving a 
D$(p-2)$-brane inside a D$p$-brane in the case of non-vanishing $b$ field, for
$p=3$. On the other hand, equations (\ref{spike}) and (\ref{spike1})
describe a tilted BIon, this time being delocalized in the direction
where the D-string lies along $\sigma^1$ direction, (\ref{spike2}).

We can study two different T-duality transformations,
since there are two inequivalent world volume directions. Let us study
T-duality along direction $\sigma^1$. Proceeding as before, the BPS equations 
in the T-dual configuration turn out to be
\bea
\CF' &=& \tan \A \\
\6_0 z' &=& -G'_{03}  \\
\CF'_{0\8a} &=& -\6_{\8a} y'
\eea
which correspond to a non-threshold D0-D2 bound state with some (tilted)
BIon ending on it, boosted in the compact direction.

Instead, we could have T-dualized along the $\sigma^3$ direction. The 
truncation of the BPS equations is
given by
\bea
\CF'_{01} &=& 0~ =~ \6_1 y' \\
\6_0 z' &=& -G'_{03} \\
\CF'_{02} &=& -\6_2 y' \\
\6_2 z' + G'_{23} + \6_2 y' G'_{3y} &=& \tan\A
\eea
which describe a non-threshold D2-D2 bound state with some (tilted) BIon 
ending on it, again, boosted in the compact direction.

\section{Arbitrary bosonic background}
\indent

In previous sections, we showed that in constant $G_{mn}$ and
$b_{mn}$ backgrounds, bosonic configurations satisfying the kappa
symmetry preserving condition (\ref{spc}) are mapped under T-duality
to the corresponding bosonic configurations in the T-dual picture.
We would like to extend that proof for an arbitrary bosonic background.

It is well known that D-branes are kappa symmetric whenever the
background satisfies the superspace constraints \cite{BT,Cw}. The
structure of kappa symmetry transformations is always given by
the requirements
\bea
\delta_\kappa Z^M E^{\uaa}_M &=& 0 \\
\delta_\kappa Z^M E_M^{\ua} &=& {1\over 2}\left(1+\G_\kappa\right)^{\ua}
_{\ub}\kappa^{\ub} \, .
\label{gkappa}
\eea
where $E_M^{\underline{A}}$ are the different components of the
supervielbeins, which should be thought of as power expansions in the
fermionic $\theta$ fields and
\begin{eqnarray}
\Gamma_\kappa &=& \frac{1}{\sqrt{-det\,(\CG + \CF)}}\sum_{l=0}
\gamma_{(2l+1)}\,\Gamma_{11}^{l+1}\wedge e^{\CF} \quad (type\, IIA) \\
\Gamma_{\kappa} &=& \frac{1}{\sqrt{-det\,(\CG + \CF)}}\sum_{l=0}
\gamma_{(2l)}\tau_3^l \wedge e^{\CF}\,i\tau_2 \quad (type\, IIB)\,,
\end{eqnarray}
where
\begin{eqnarray}
\gamma_{(1)}&=& d\sigma^\mu\gamma_\mu = d\sigma^\mu \6_\mu Z^M 
E_M^{\underline{a}}\Gamma_{\underline{a}} \\
%\CF &=& F - \frac12dZ^M E_M^{\underline{A}}\wedge dZ^N E_N^{\underline{B}}
%B_{\underline{A}\underline{B}}\,.
\CF &=& F - \frac12dZ^M E_M^{\underline{A}}\wedge dZ^N E_N^{\underline{B}}
B_{\underline{A}\underline{B}}\,.
\end{eqnarray}

It is nevertheless true that the condition for any bosonic configuration
$(\theta=0)$ to preserve some supersymmetry is still given by
$\G_\kappa \ep = \ep$. The reason is that when studying the $\theta=0$
limit of the supervielbein components \cite{fkap},
\begin{equation}
E_m^{\uaa} |_{\theta=0} = e_m^{\uaa}(x^n) \, ,~~~~ \, 
E_m^{\ua}|_{\theta=0} = 0\, ,~~~~ 
\, 
B_{mn}|_{\theta=0} = b_{mn}(x^n)
\end{equation}
so that 
\begin{equation}
\delta_\kappa \theta^{\A}e_{\A}^{\ua} = {1\over 2}
\left(1+\G_\kappa|_{\theta=0}\right)^{\ua}_{\ub}\kappa^{\ub}\, ,
\end{equation} 
which determines the universal condition
\begin{equation}
\G_\kappa|_{\theta=0}\ep = \ep \, ,
\label{spc1}
\end{equation}
$\ep$ being the Killing spinor of the corresponding bosonic supergravity
background. 

$\G_\kappa|_{\theta=0}$ depends on the background geometry, but since the 
T-duality rules for the bosonic sector of the supergravity fields are known 
\cite{bho}
\begin{eqnarray}
G_{zz} & = & 1/G'_{\7z\7z} \nonumber \\
b_{nz} & = & -G'_{n\7z}/G'_{\7z\7z} \nonumber \\
G_{nz} & = & -b'_{n\7z}/G'_{\7z\7z} \nonumber \\
G_{mn} & = & G'_{mn} - (G'_{m\7z}G'_{n\7z}-b'_{m\7z}b'_{n\7z})/G'_{\7z\7z} 
\nonumber \\
b_{mn} & = & b'_{mn} - (b'_{m\7z}G'_{n\7z}-b'_{n\7z}G'_{m\7z})/G'_{\7z\7z}\,
\label{Tdu}
\end{eqnarray}
one can indeed compute the behaviour of $\G_\kappa|_{\theta=0}$ under 
T-duality, as we did previously \footnote{In the following we are explicitly
using the parametrisation $\eta=a_2=-a_1=1$ and $\Gamma_m~^n=\delta_m~^n$.}. 
Using the same notation as in previous sections
\begin{equation}
\gamma_{(1)}|_{\theta=0}= dx^m e_m~^{\underline{a}}\Gamma_{\underline{a}} 
=\slPi= \8\bslPi~+~\G_{\uz} D\rho,
\end{equation}
where
\bea
\8\bslPi&\equiv&\G_{\8\uaa}\bd x^m e^{\8\uaa}_m(x^n)
\label{firstpi1} \\
D\rho&\equiv&\Pi^\uz~=~\lambda d\rho +\bd x^m e^{\uz}_m(x^n).
\eea

Under T-duality, it can be checked that
\begin{eqnarray}
\8\bslPi &=& \8\bslPi' \\
\CF &=& \CF' + D\rho\bPi^{'\uz}\,,
\label{gentdu}
\end{eqnarray} 
where
\begin{equation}
\bPi^{'\uz} = \lambda'\bd z' +\bd x'^m e_m^{'\uz}\,.
\end{equation}

Once (\ref{gentdu}) is known, it is straightforward to extend
the techniques developed in appendix B and C to show that
any bosonic configuration solving (\ref{spc1}) in type IIA,
is mapped to the corresponding T-dual one satisfying
\begin{equation}
\Gamma_{\kappa'}|_{\theta'=0}\epsilon'=\epsilon'\,,
\end{equation}
where the relation among Killing spinors is given by
\begin{equation}
\ep_1=\ep'_2, \quad , \quad 
%{\G_{\uz}e^{\uz}_z\over\sqrt{G_{zz}}}\ep_2 =\ep'_1,~~~~~~\to~~~~~~~
{\G_{\uz}}\ep_2 =\ep'_1,
\end{equation}
which is consistent with the transformations found in \cite{new},
since we have used a Lorentz (gauge) rotation to set $e_z^{\underline{a}}=0$.

\section{Discussion}
\indent

We would like to finish with some discussion concerning
possible natural extensions of the results presented in this paper. 
In particular, we will concentrate on three subjects : 
\begin{itemize}
\item
T-duality realized on D-branes coupled to an arbitrary kappa symmetric 
background.
\item Non-BPS D-branes.
\item Non-abelian D-branes.
\end{itemize}

{\em Arbitrary kappa symmetric backgrounds.}
There has been some recent interest in the open problem related to
T-duality in curved kappa symmetric backgrounds \cite{new}. When trying 
to extend our approach to this general case, it seems rather natural to 
demand the relations
\begin{equation}
\theta^\A_1 E_\A^{\ua} = a_2 \theta'^\A_2 E'^{\ua}_\A \quad , \quad
\theta^\A_2 E_\A^{\ua} = -a_1 \left(\G_{\uz}\right)^{\ua}_{\ub}
E'^{\ub}_{\A}\theta'^{\A}_1.
\label{ansatz}
\end{equation} 
Equations (\ref{ansatz}) deserve several remarks. First of all, they
are reminiscent of the extension of the kappa symmetry transformations from 
the SuperPoincar\'e case to the arbitrary kappa symmetric background. 
Secondly, it is not clear which is the solution to them,
that is, $\theta'^\A_{1|2} = f^\A_{-|+} (\theta^\B)$, since the supervielbeins
appearing in both sides of them admit an expansion in the corresponding
fermionic fields. Finally, the mapping between fermionic
fields will be non-constant in general, so that when computing the
T-duality transformation of the operators coupling to derivatives
of these fermionic fields, they will involve components of the spin
connection.

Irrespectively of which is the real solution, the latter should certainly
satisfy some constraints. First of all, it should be such that the
T-duality rules for the closed string sector must map the supergravity
constraints of type IIA to the ones of type IIB. This is equivalent to
map the D-brane effective action and its kappa symmetry structure in
type IIA to the corresponding ones in type IIB. In other words, the
mapping should be {\em T-duality covariant} and satisfy
\bea
& \delta_\kappa Z^M E^{\uaa}_M = 0 \,\longrightarrow \, 
\delta_{\kappa'} Z'^M E'^{\uaa}_M = 0 & \\
& \delta_\kappa Z^M E_M^{\ua} = {1\over 2}\left(1+\G_\kappa\right)^{\ua}
_{\ub}\kappa^{\ub} \, \longrightarrow \, 
\delta_{\kappa'} Z'^M E'^{\ua}_M = {1\over 2}\left(1+\G'_{\kappa'}\right)^{\ua}
_{\ub}\kappa'^{\ub} \, .
\eea

{\em Non-BPS D-brane effective actions.} It has recently been argued
that the effective action describing a non-BPS D-brane probing in
SuperPoincare should be splitted into a DBI term \cite{sen}
\begin{equation}
S_{non-BPS}=-\int\,d^{p+1}\sigma \,\sqrt{-\,det\left(\CG + \CF\right)}
\,f(T,\partial_\mu T, \ldots {\tilde \CG}^{\mu\nu}_S,{\tilde \CG}^{\mu\nu}_A)
\label{nonBPS1}
\end{equation}
plus a WZ term describing the coupling of the tachyonic scalar field $T$
to the R-R sector \cite{WZ},
\begin{equation}
S_{WZ} = \int_{M_{p+2}} {\cal C}\wedge\,dT\,\wedge e^{\CF}\, .
\label{nonBPS2}
\end{equation}
Due to the scalar character of the tachyonic field, it is natural to
extend the functional truncation $(\6_\rho \phi^{\8{\i}}=0)$ to it,
$\6_\rho T =0$. In this way, it is straightforward, using our
previous analysis, to check that WZ terms (\ref{nonBPS2}) are indeed 
T-duality covariant, as they should be. 
When being concerned about the T-duality
properties of (\ref{nonBPS1}), we do appreciate an important characteristic
of the T-duality covariance requirement. Indeed, T-duality covariance does
not fix the effective dynamics of the open string sector by itself, it
just constraints it. For example, (\ref{nonBPS1}) must be T-duality
covariant, which means that 
$f(T,\partial_\mu T, \ldots {\tilde \CG}^{\mu\nu}_S,{\tilde \CG}^{\mu\nu}_A)$
is covariant, since the usual DBI square root is. This requirement does
not fix $f$. For instance, we can not distinguish between \footnote{JS
would like to thank Eduardo Eyras for a discussion concerning this point.}
\begin{equation}
\sqrt{-\,det\left(\CG_{\mu\nu} + \CF_{\mu\nu}+ \6_\mu T \6_\nu T\right)}
\end{equation}
and
\begin{equation}
\sqrt{-\,det\left(\CG + \CF\right)}\,\sum_n a_n \left(\7\CG^{\mu\nu}_S\6_\mu T 
\6_\nu T\right)^n
\end{equation}
for arbitrary constant coefficients $a_n$, both being T-duality covariant
due to the covariance of 
$\7\CG^{\mu\nu}_S=(\CG + \CF)^{-1(\mu\nu)}$. See \cite{eduardo} for
a discussion of T-duality properties of non-BPS D-brane effective actions.

{\em Non-abelian D-branes.}
In \cite{nonab}, the approach followed in this paper was used to determine
the effect of non-trivial commutators among scalar fields in the non-abelian
bosonic generalization of the DBI action. The main idea there was to assume 
that the trace over the $U(N)$ gauge group indices was the symmetrized one 
(again T-duality does not fix this possibility) and
study the dimensional reduction of the D9-brane field theory where
world volume diffeomorphisms had been gauge fixed (since no covariant
version is known for non-abelian D-brane effective actions). 

In the following, we will briefly comment on the extension of that result
to non-abelian SuperD-branes propagating in SuperPoincar\'e. As in 
\cite{nonab},
we will assume a symmetrized prescription for the trace and replace all
partial derivatives by covariant derivatives. Since the new action includes 
fermions, one must also gauge fix kappa symmetry. 
Following \cite{Shgf}, we choose
\begin{equation}
\theta_1=0 \quad , \quad \theta_2=\Lam
\end{equation}
ensuring the vanishing of the WZ term, so that we concentrate on the
DBI term of the effective action. The components of the tensor
\begin{equation}
E_{\mu\nu} = \Pi^m_\mu\Pi^n_\nu\eta_{mn} +\CF_{\mu\nu}\, 
\end{equation}
can be written after the gauge fixing as
\begin{eqnarray}
E_{\hmu\hnu} & = & \eta_{\hmu\hnu} -2\bLam\G_{\hmu}D_{\hnu}\Lam +
F_{\hmu\hnu} + (\bLam\G^mD_{\hmu}\Lam)(\bLam\G^nD_{\hnu}\Lam)\eta_{mn} \\
E_{\hmu}^{~i} & = & -2i\bLam\G_{\hmu}[x^i,\Lam] + i(\bLam\G^mD_{\hmu}\Lam)
(\bLam\G^n[x^i,\Lam])\eta_{mn} + D_{\hmu}x^i \\
E^i_{~\hmu} & = & -2\bLam\G_{i}D_{\hmu}\Lam + i(\bLam\G^mD_{\hmu}\Lam)
(\bLam\G^n[x^i,\Lam])\eta_{mn} - D_{\hmu}x^i \\
E^{ij} & = & \delta^{ij} - (\bLam\G^m[x^i,\Lam])(\bLam\G^n[x^j,\Lam])\eta_{mn}
-2i\bLam\G^i[x^j,\Lam] + i[x^i,x^j]
\end{eqnarray} 
where we used the same reduction rules as those used in \cite{nonab}, with the
addition that $D_i\Lam=i[x^i,\Lam]$, and we splitted the initial world volume 
directions $\{\mu,\nu\}$ into T-dual world volume ones $\{\hmu,\hnu\}$ and 
transverse directions denoted by scalars $\{x^i\}$.

By introducing the matrix
\begin{equation}
Q^i_k = \delta^i_k + i[x^i,x^j]\delta_{jk} -2i\bLam\G^i[x^j,\Lam]
\delta_{jk} - (\bLam\G^m[x^i,\Lam])(\bLam\G^n[x^j,\Lam])\eta_{mn}
\delta_{jk}
\end{equation}
we can rewrite $E^{ji}$ and its inverse $E_{ik}$ as 
\begin{eqnarray}
E^{ji}&=&\delta^{ji} + (Q^j_k -\delta^j_k)\delta^{ki}=Q^j_k\delta^{ki} \\
E_{ik}&=&\delta_{il}(Q^{-1})^l_k.
\label{def1}
\end{eqnarray}
In this way, we can now compute the determinant of the ten dimensional 
original matrix (notice that $det \,E^{ij}= det \,Q^i_k$) :
\begin{eqnarray}
det\,(\CG + \CF) & = & det\,A \, det \,Q \\
A_{\hmu\hnu} & = & E_{\hmu\hnu} - E_{\hmu}^{~i} E_{ik} E^k_{~\hnu} \, ,
\end{eqnarray}
thus generalizing the result presented in \cite{nonab}.

\section*{Acknowledgements}
The authors would like to thank Joaquim Gomis for valuable discussions.
JS is supported by a fellowship from Comissionat per a Universitats i Recerca
de la Generalitat de Catalunya. This work was supported in part by
AEN98-0431 (CICYT), GC 1998SGR (CIRIT).

\appendix

\section{Proof of T-duality covariance}\label{appA}

\indent

In this appendix, we will analyze the constraints derived from
requiring T-duality covariance of the DBI term
\begin{equation} 
-T_p\,\int d^{p+1}\sigma \,\sqrt{-det\left(\CG+\CF\right)}
\limit{\longrightarrow}{\CT_{\parallel}}
-T'_{p-1}\,\int d^{p}\sigma \,\sqrt{-det\left(\CG'+\CF'\right)}\,.
\end{equation}
For this mapping to be satisfied, it is sufficient to hold
\begin{equation}
T'_{p-1}~=~T_p R,~~~~~~~R~\equiv~\int d\rho \sqrt{G_{zz}}
\label{cond1}
\end{equation}
which is derived from operators involving no derivative of the dynamical
fields $\{\phi^{'\8{\i}}\}$, and
\bea
\CG_{\hmu\hnu}+\CF_{\hmu\hnu} - {1\over \CG_{\rho\rho}}
(\CG_{\hmu\rho}+\CF_{\hmu\rho})(\CG_{\hnu\rho}-\CF_{\hnu\rho})&=&
\CG'_{\hmu\hnu}+\CF'_{\hmu\hnu}
\label{cond2}
\eea
from operators involving such derivatives $\{\6_{\hmu}\phi^{'\8{\i}}\}$.

Equation (\ref{cond1}) gives the correct tension for the T-dual D-brane.
Notice that it is equivalent to the usual T-duality transformation
for the dilaton field
\begin{equation}
\phi^\prime = \phi -\frac{1}{2}\log\vert G_{zz}\vert 
\end{equation}
when the latter is constant. Equations (\ref{cond2}) are further split into 
their symmetric and anti-symmetric parts
\bea
\CG_{\hmu\hnu} - {1\over \CG_{\rho\rho}}(
\CG_{\hmu\rho}\CG_{\hnu\rho}-\CF_{\hmu\rho}\CF_{\hnu\rho})
&=&\CG'_{\hmu\hnu},
\label{cond2s}\\
\CF_{\hmu\hnu} + {1\over \CG_{\rho\rho}}
(\CG_{\hmu\rho}\CF_{\hnu\rho}-\CF_{\hmu\rho}\CG_{\hnu\rho})&=&
\CF'_{\hmu\hnu}.
\label{cond2a}
\eea
The induced metric on the IIA D$p$-brane is given by
\bea
\CG_{\rho\rho}&=&G_{zz},
\nn\\
\CG_{\hmu\rho}&=&\pa_\hmu Z^\hM G_{\hM z},
\nn\\
\CG_{\hmu\hnu}&=&\pa_{\hmu} Z^{\hM} \pa_{\hnu} Z^{\hN}~(-)^{MN} G_{\hM\hN},
\eea
where we took into account the conditions (\ref{pardif})-(\ref{cons})
defining $\CT_{\parallel}$, while the one on the IIB \Dpm-brane is just
\bea
\CG'_{\hmu\hnu}&=&
\pa_{\hmu} z'\pa_{\hnu} z' G'_{zz}~+~\pa_{(\hmu} Z'^{\hM} \pa_{\hnu)} z'
G'_{\hM z}~+~\pa_{\hmu} Z'^{\hM} \pa_{\hnu} Z'^{\hN}~(-)^{MN}G'_{\hM\hN}.
\eea
$(-)^{MN}=-1$ when both M and N are odd, $(-)^{MN}=1$ for others.
On the other hand, the components of the gauge invariant tensor $\CF/(\CF')$ 
on the IIA/(IIB) D-branes can be decomposed as 
\bea
\CF_{\hmu\hnu}&=&
\pa_{[\hmu}V_{\hnu]}-\frac12\pa_{[\hmu} Z^\hM\pa_{\hnu]} Z^\hN~B_{\hM\hN}
\nn\\
\CF_{\hmu\rho}&=&\pa_{\hmu}V_{\rho}+\pa_{\hmu} Z^\hN~B_{z\hN}
\\
\CF'_{\hmu\hnu}&=&
\pa_{[\hmu}V_{\hnu]}-\frac12\pa_{[\hmu} Z'^\hM\pa_{\hnu]} Z'^\hN~B'_{\hM\hN}~-~
\pa_{[\hmu} z'\pa_{\hnu]} Z'^\hN~B'_{z\hN}.
\eea
\vs

Using these decompositions in \bref{cond2s}, and matching the coefficients of
the different independent operators $\{\6_{\hmu}\phi^{'\8{\i}}
\6_{\hnu}\phi^{'\8{\j}}\}$ appearing in both sides, we find
\bea
 {{\h^2}\over G_{zz}}&=&G'_{zz},~~~~
\label{eq1a} \\
{{-\h}\over G_{zz}} B_{\hN z}&=&
\G_\hN^{~\hM} G'_{z\hM}
\label{eq2a} \\
G_{\hM\hN} - {1\over G_{zz}}
[ G_{\hM z}G_{\hN z}-
B_{z\hM}B_{z\hN}]&=&(-)^{(M+M')N}
\G_{\hM}^{~\hM'}\G_{\hN}^{~\hN'}~G'_{\hM'\hN'}.
\label{eq3a}
\eea
Proceeding in the same way with \bref{cond2a}, we obtain
\bea
{{-\h}\over G_{zz}}G_{\hM z}&=&\G_\hM^{~\hN}~B'_{\hN z}
\label{eq4a}
\eea
\bea
&& B_{\hM\hN}-{1\over G_{zz}}
[(-)^{MN}\{G_{\hM z}B_{z\hN}~-~B_{z\hM}G_{\hN z}\}]~=~
(-)^{(M+M')N}
\G_{\hM}^{~\hM'}\G_{\hN}^{~\hN'}~B'_{\hM'\hN'}.
\nn\\
\label{eq5a}
\eea
Equations \bref{eq1a}-\bref{eq5a} are the set of constraints derived from
the T-duality covariance requirement. They can be interpreted
as the generalization of the usual bosonic T-duality rules for the
kind of superfields we are considering along the whole paper.

In the following, we will start analyzing equations \bref{eq1a}-\bref{eq5a}.
Before doing so, we must identify the different components of the
superfields appearing in them.  We can read the components of the superspace 
metric $G_{MN}={e_M}^{\underline{a}}{e_N}^{\underline{b}}
\eta_{\underline{a}\underline{b}}$ in type IIA from (\ref{susyinv})-
(\ref{susyinv1})
\bea
G_{mn}&=&e_m^{~\uaa}e_n^{~\ubb}\h_{\uaa\ubb},~~~~~~~~~~~~~~~~~~~~~~~~~~~~
\nn\\
G_{i\A,m}&=&(\Ttb_i\G_\uaa )_\ua e_\A^{~~\ua}~e_m^{~~\uaa}\equiv~
(\Ttb_i\G_\uaa )_\A~e_m^{~~\uaa},~~~~~~~~~~~~~~~~~~~~~~
\nn\\
G_{i\A,j\B}&=&(\Ttb_i\G_\uaa )_\A (\Ttb_j\G^\uaa )_\B,~~~~~~
\eea
and those of the NS-NS superfield $B_{MN}$  from (\ref{twoform})
\bea
B_{mn}&=&b_{mn},
\nn\\
B_{i\A,m}&=&-(-1)^i(\Ttb_i\G_\uaa )_\A~e_m^{~~\uaa},
\nn\\
B_{i\A,j\B}&=&(-1)^i(\Ttb_i\G_\uaa )_\A(\Ttb_j\G^\uaa )_\B.
\eea
In type IIB, $G'_{MN}$ is decomposed as
\bea
G'_{mn}&=&e_m^{'~\uaa}e_n^{'~\ubb}\h_{\uaa\ubb},~~~~~~~~~~~~~~~~~~~~~~~~~~~~
\nn\\
G'_{i\A,m}&=&(\Ttb'_i\G_\uaa )_\ua e_\A^{'~~\ua}~e_m^{'~~\uaa}\equiv~
(\Ttb'_i\G_\uaa )_\A~e_m^{'~~\uaa},~~~~~~~~~~~~~~~~~~~~~~
\nn\\
G'_{i\A,j\B}&=&(\Ttb'_i\G_\uaa )_\A (\Ttb'_j\G^\uaa )_\B,~~~~~~
\eea
while the NS-NS superfield $B'_{MN}$  as
\bea
B'_{mn}&=&b'_{mn},
\nn\\
B'_{i\A,m}&=&(-1)^i(\Ttb'_i\G_\uaa )_\A~e_m^{'~~\uaa},
\nn\\
B'_{i\A,j\B}&=&-(-1)^i(\Ttb'_i\G_\uaa )_\A(\Ttb'_j\G^\uaa )_\B.
\eea

It is always possible to make a local $SO(1,9)$ rotation in both IIA/IIB 
tangent spaces to set
\bea
e_z^{~~\uaa}&=&\Lam~\D_{\uz}^{~~\uaa},~~~~~~
e_z^{'~~\uaa}~=~\Lam'~\D_{\uz}^{~~\uaa},
\eea
where $\Lam$ and $\Lam'$ are constants such that 
\bea
G_{zz}&=&\Lam^2,~~~~~~G'_{zz}~=~{\Lam'}^2.
\eea
Equation \bref{eq1a} becomes
\bea
\h^2&=&\Lam^2~{\Lam'}^2\,,
\label{radius}
\eea
which is the analogue of the usual T-duality rules relating the radius
of the original circle with the radius of the T-dual circle, 
$G'_{z'z'}G_{zz}=1$.

\noindent
Equations \bref{eq2a} and \bref{eq3a} allow us to set some elements
of $\G_M^{~N}$ to zero
\bea
\G^{~j\B}_{m}~=~\G^{~m}_{j\B}~=~\G^{~1\A}_{1\B}~=~\G^{~2\A}_{2\B}&=&0
\eea
and
\bea
s~\frac{b_{mz}}{\Lam}&=&\G_m^{~n}~e'_{n\uz},~~~~~~~~~
s~e_{m\uz}~=~\G_m^{~n}~\frac{b'_{nz}}{\Lam'},
\label{eqbed1}\eea
\bea
s~(\Ttb_i\G_\uz)_\A&=&-(-1)^i~{\G_{i\A}}^{j\B}~(\Ttb'_j\G_\uz)_\B~,
\label{eq417a}
\eea
where $s\equiv~\frac{-\h}{\Lam\Lam'}=\pm 1$ is a signature.
\noindent
Assuming IIB spinors $\T'_j$ have positive chirality, we can take
\bea
&&\Ttb_1~=~a_2 \Ttb'_2,~~~~~~\Ttb_2~=~a_1 \Ttb'_1\G_\uz.
\label{ttrel}
\eea
and
\bea
s~a_2~ e_\A^{~~\ua}&=&
e_\B^{'~~\ua}~{\G_{1\A}}^{2\B},~~~~~~~~~
-s~a_1 ~(\G_\uz)^\ub_{~~\ua} e_\A^{~~\ua}~=~
      e_\B^{'~~\ub}~{\G_{2\A}}^{1\B}.
\label{eq418b}
\eea
It follows, in addition to \bref{eq417a}, for $\8\uaa\neq \uz$ components that
\bea
s~(\Ttb_i\G_{\8\uaa})_\A &=&
(\Ttb'_j\G_{\8\uaa})_\B~{\G_{i\A}}^{j\B}.
\label{eq417d}
\eea
From the equation \bref{eq4a} we get
\bea
e_{m}^{~~\8\uaa}&=&s~
\G^{~~m'}_{m}~e_{m'}^{'~~\8\uaa}.
\label{eerel1}
\eea
Finally \bref{eq5a} requires
\bea
b_{mn}~-~\frac{b_{[mz}}{\Lam}~e_{n]\uz}&=&
\G^{~~m'}_{m}~\G^{~~n'}_{n}~b'_{m'n'}.
\label{eqbmn1}
\eea

\noindent
Thus, as we claimed in the introduction, T-duality covariance of the DBI
action fixes the chirality change mapping among spinor fields (\ref{ttrel})
up to constant factors $(a_1,a_2)$ from the fermionic components of the
background superfields, and reproduces the well known transformations for their
bosonic components (\ref{radius}), (\ref{eqbed1}), (\ref{eerel1}) and
(\ref{eqbmn1}). 

Having solved such T-duality constraints, we will next study the
T-duality properties of the different supersymmetric invariant forms
defined on D-branes. Let us consider $\slPi=\Gamma_{\underline{a}}
\Pi^{\underline{a}}$ in type IIA. It can be splitted in terms of
\bea
\slPi&=&\8\bslPi~+~\G_{\uz} D\rho,
\eea
where
\bea
\8\bslPi&\equiv&\G_{\8\uaa}(\bd x^m e^{~\8\uaa}_m+\Ttb\G^{\8\uaa} \bd \7\T)
\label{firstpi} \\
D\rho&\equiv&\Pi^\uz~=~\lambda d\rho +\bd x^m e^{~\uz}_m+ \Ttb\G^\uz \bd\7\T.
\eea
Analogously, in type IIB,
\bea
\slPi'
&=&\8\bslPi'~+~\G_{\uz} \bPi^{'\uz},
\eea
where
\bea
\8\bslPi'&\equiv&\G_{\8\uaa}(\bd x^{'m} e^{'~\8\uaa}_m+\Ttb'\G^{\8\uaa} 
\bd \7\T')
\nn\\
 \bPi^{'\uz}&\equiv&
(\lambda' d z' + \bd x^{'m} e^{'~~\uz}_m+ \Ttb'\G^\uz \bd\7\T').
\label{bPid}
\eea
We can write $~\8\bslPi$ in terms of type IIB variables by inserting
(\ref{ttrel}), (\ref{eq417d}) and (\ref{eerel1}) into (\ref{firstpi})
\bea
\8\bslPi\equiv\G_{\8\uaa}(\bd x^m e^{~\8\uaa}_m+\Ttb\G^{\8\uaa} \bd \7\T)=
\G_{\8\uaa}(\bd x^m (s \G_m^{~m'}e'^{~\8\uaa}_{m'})+(
a_2^2~\Ttb'_2\G^{\8\uaa} \bd \7\T'_2+a_1^2~\Ttb'_1\G^{\8\uaa} \bd \7\T'_1)).
\eea
Thus, by choosing
\bea
s~=~+1,~~~~~~a_1^2~=~1,~~~~~~a_2^2~=~1
\label{a1a2}
\eea
we can identify both one forms
\bea
\8\bslPi&=&
\G_{\8\uaa}(\bd x^{'m}e'^{~\8\uaa}_{m}+\Ttb'\G^{\8\uaa} \bd \7\T')~=~\8\bslPi'.
\label{pirel}
\eea
The latter equation is telling us that the supersymmetric invariant
one form in nine dimensions is T-duality covariant.
Furthermore, when using \bref{a1a2} in (\ref{ttrel}) and (\ref{eq417d}),
it follows that
\bea
\Ttb\G^{\8\uaa} \bd \7\T&=&+~\Ttb'\G^{\8\uaa} \bd \7\T',~~~~~~~~~
\Ttb\G_{11}\G^{\8\uaa} \bd \7\T~=~+~\Ttb'\tau_3\G^{\8\uaa} \bd \7\T',
\\
\Ttb\G^{\uz} \bd \7\T&=&-~\Ttb'\tau_3\G^{\uz} \bd \7\T',~~~~~~~
\Ttb\G_{11}\G^{\uz} \bd \7\T~=~-~\Ttb'\G^{\uz} \bd \7\T',
\eea
which are the form version of the identities (\ref{conditions}),
and
\bea
D\rho&=&\lambda d\rho +\bd x^m e^{~\uz}_m+ \Ttb\G^z \bd\7\T ~=~
\lambda d\rho~ +~\bd x^{'m}\frac{b'_{m\uz}}{\Lam'}~-~ 
\Ttb'\tau_3\G^z \bd\7\T'.
\eea
Concerning the supersymmetric invariant two form $\CF$
\bea
\CF&=&dV~+~(\Ttb\ggt\G_md\7\T)(d\7 x^m~+~
\frac12\Ttb\G^m d\7\T)-{1\over 2}dx^mdx^nb_{mn}\,,
\eea
it can be written in terms of type IIB variables as 
\bea
\CF&=&\CF'~+~D\rho~\bPi'^{\uz},
\label{FFdrel}
\eea
where
\bea
\CF'&\equiv&
\bd {\bV}-{1\over 2}dx^{'m}dx^{'n}b'_{mn} 
 +(\Ttb'\tau_3\G_{\uaa}\bd \7\T')(\bd x^{'m}e'^{~\uaa}_m~+~\frac12\Ttb'
\G^{\uaa}\bd\7\T').
\eea

It is remarkable that all terms $i_{\6_\rho}\CF$ appearing in the
decomposition of the supersymmetric invariant form $\CF$ under the double
dimensional ansatz, $\CF=\CF^- + d\rho\wedge i_{\6_\rho}\CF$, can be written
as $\bPi^{'\uz}$, the supersymmetric invariant one form along the T-dual 
circle. Furthermore, all the dependence of $d\rho$ in supersymmetric
invariant forms $\slPi$ and $\CF$ is through $D\rho=\Pi^{\uz}$, the supersymmetric
one form along the original circle. Thus, what T-duality does is to exchange
both forms $\Pi^{\uz} \longleftrightarrow\bPi^{'\uz}$. This is the 
supersymmetric generalization of the corresponding phenomena observed in
the bosonic case \cite{joan}, whose relevance may become more clear in the
discussion of the T-duality transformation of the WZ term in appendix B.

%%%%%%%%%%%%%%%%%%%%%%%%%%%%%%%%%%%%%%%%%%%%%%%%%%%%%

\section{T-Duality transformation of WZ term}
\indent
In this appendix we prove that WZ terms of type IIA SuperD-branes are mapped
to WZ terms of type IIB SuperD-branes under T-duality, using the results
obtained in appendix A. WZ terms of IIA D-branes are obtained from 
\bea
dL^{WZ}_A~=~-~T_p~\ba E~\CC_A~E~e^\CF,~~~~~~~~
\CC_A(\slPi)~=~\sum_{\l=0}(\ggt)^{\l+1}\frac{\slPi^{2\l}}{(2\l!)},
\label{IIAWZ}
\eea
where 
\bea
E^\ua~=~d\7\T^{\ua}~=~d\T^\A e_\A^{~~\ua},~~~~~~~
\ba E_\ub~=~d\7\T^{\ua}C_{\ua\ub}
,~~~~~~~~
\slPi~=~\G_\uaa(d\7x^\uaa+\ba{\7\T}\G^\uaa d\7\T).
\eea
The latter show that $dL^{WZ}_A$ just depends on supersymmetric invariant 
forms, whose T-duality properties were determined in appendix \ref{appA}.
In particular, from (\ref{FFdrel}) and using $D\rho\wedge D\rho=0$,
\bea
e^\CF&=&\sum_{n=0}\frac{\CF^n}{n!}~=~\sum_{n=0}\frac{(\CF'+D\rho ~\bPi'^\uz)
^n}{n!}~=~
(1~+~(D\rho ~\bPi'^\uz))~e^{\CF'}.
\label{B7}
\eea
Next, using (\ref{eq417d}) and (\ref{pirel}) besides that $\8\bslPi$ and 
$\G^\uz D\rho $ are commuting one forms,
\bea
\ba E~\CC_A~E
&=&\bd\Ttb~\sum_{\l=0}(\ggt)^{\l+1}\frac{\slPi^{2\l}}{(2\l!)}~\bd\Tt
~=~
\bd\Ttb~\sum_{\l=0}(\ggt)^{\l+1}\frac{(\8\bslPi+\G^\uz D\rho)^{2\l}}{(2\l!)}~
\bd\Tt
\nn\\
&=&\bd\Ttb_2\frac{\8\bslPi^{2\l}}{(2\l)!}\bd\Tt_1~+~
   \bd\Ttb_1(-1)^{\l+1}\frac{\8\bslPi^{2\l}}{(2\l)!}\bd\Tt_2
\nn\\&&~+~
\bd\Ttb_2(\G^\uz D\rho)\frac{\8\bslPi^{2\l+1}}{(2\l+1)!}\bd\Tt_1~+~
   \bd\Ttb_1(\G^\uz D\rho)(-1)^{\l}\frac{\8\bslPi^{2\l+1}}{(2\l+1)!}\bd\Tt_2
\nn\\
&=&a_1a_2\{\bd\Ttb_1'\G_\uz\frac{\8\bslPi^{2\l}}{(2\l)!}\bd\Tt_2'~+~
   \bd\Ttb_2'(-1)^{\l+1}\frac{\8\bslPi^{2\l}}{(2\l)!}\G_\uz\bd\Tt_1'
\nn\\&&~+~
\bd\Ttb_1' D\rho\frac{\8\bslPi^{2\l+1}}{(2\l+1)!}\bd\Tt_2'~+~
   \bd\Ttb_2' D\rho(-1)^{\l}\frac{\8\bslPi^{2\l+1}}{(2\l+1)!}\bd\Tt_1'\}
\nn\\
&=&a_1a_2\{\bd\Ttb'\frac{\8\bslPi^{2\l}}{(2\l)!}\G_\uz\tau_3^\l\tau_1\bd\Tt'~
+~\bd\Ttb' D\rho\frac{\8\bslPi^{2\l+1}}{(2\l+1)!}\tau_3^\l\tau_1\bd\Tt'\}.
\label{B8}
\eea
Joining (\ref{B7}) and (\ref{B8})
\bea
dL^{WZ}_A&=&-~T_p~a_1a_2~
\bd\Ttb'[\sum_{\l=0}\frac{\8\bslPi^{'2\l}}{(2\l!)}\G^\uz\tau_3^\l\tau_1
+( D\rho)\sum_{\l=0}\frac{\8\bslPi^{'2\l+1}}{(2\l+1)!}\tau_3^\l\tau_1]
\bd\Tt'
(1+( D\rho~\bPi'^z))e^{\CF'}
\nn\\
&=&-a_1a_2T_p~( D\rho)[~
\bd\Ttb'~\sum_{\l=0}\frac{(\slPi')^{2\l+1}}{(2\l+1)!}
\tau_3^\l\tau_1~\bd\Tt'~]~e^{\CF'}~+~...
\nn\\
&=&a_1a_2T_p~[~\bd\Ttb'~\CS_B(\slPi')\tau_1~\bd\Tt'~]~e^{\CF'}~(\Lam~d\rho)
~+~...,
\label{dWZA}
\eea
where dots stand for terms not depending on $d\rho$. 

{}From $dL^{WZ}_A$ in \bref{dWZA} we can find the WZ Lagrangian
$\CL^{WZ}_A$, written in terms of IIB variables, 
by taking the p+1 form part of $L^{WZ}_A$  on $\s^\mu$. 
IIB (p-1) brane WZ term will be obtained by 
integrating it over $\rho$. 
It means that only the coefficient of $ ~d\rho~$
in \bref{dWZA}  contributes to $L^{WZ}_B$.
The coefficient of $ ~d\rho$ in \bref{dWZA} 
gives  $dL^{WZ}_B$
\bea
dL^{WZ}_B&=&-T'_{p-1}~[~\ba E'~\CS_B(\slPi')\tau_1~E'~]~e^{\CF'},~~~~~~~~~~
\eea
if $~a_1a_2~=-1~$, where $T'_{p-1}$ is given in \bref{cond1}.

%%%%%%%%%%%%%%%%%%%%%%%%%%%%%%%%%%%%%%%%%%%%%%%%%%%%%%%%%%%%%%%

\section{Kappa symmetry}\label{kappaproof}
\indent
In this appendix we prove that the infinitesimal kappa symmetry transformation
$\delta_\kappa \theta$ in type IIA is mapped to $\delta_{\kappa'}\theta'$
in type IIB as claimed in (\ref{k1}). The kappa symmetry transformations for 
type IIB spinors are obtained from those of IIA using \bref{ttrel}
\bea
a_2~\D\Ttb_2'&=&\D\Ttb_1~=~\D\Ttb\G_-~=~\kb(1-\gam^{(p)})\G_-~
\label{dtd2} \\
a_1~\D\Ttb_1'&=&\D\Ttb_2\G^\uz~=~\D\Ttb\G_+~\G^\uz~=~
\D\Ttb\G^\uz\G_-~=~\kb(1-\gam^{(p)})\G^\uz\G_-.
\label{dtd1}
\eea

\noindent
First terms in the right hand side of \bref{dtd2} and \bref{dtd1} are
\bea
\kb\G_-&=&\kb_1~\equiv~a_2~\kb_2', \\
\kb\G^\uz\G_-&=&\kb\G_+ \G^\uz  ~=~
\kb_2\G^\uz~\equiv~a_1~\kb'_1,
\eea
where we assumed that kappa symmetry parameters $\kappa_j$ have the same 
T-duality transformation as the dynamical fields $\T$
 in \bref{ttrel},
\bea
\kb_1~=~a_2~\kb_2',~~~~~~\kb_2~=~a_1~\kb_1'~\G^\uz.
\eea
The second term in the right hand side of \bref{dtd2} is
\bea
-\kb(\gam^{(p)})\G_-
&=&\frac{-\kb}{\sqrt{-det(\CG+\CF)}}[\CS_A(\slPi)e^\CF]_{p+1}
\G_-
\nn\\&=&
\frac{-\kb}{\sqrt{-det(\CG+\CF)}}[\sum_{\l=0}\frac{\slPi^{2\l+1}}
{(2\l+1)!}e^\CF]_{p+1 }\G_-
\nn \\
&=&\frac{-\kb}{\sqrt{-det(\CG+\CF)}}[\sum_{\l=0}
\frac{(\8\bslPi+\G^\uz D\rho)^{2\l+1}}
{(2\l+1)!}(1+ D\rho\bPi'_z)e^{\CF'}]_{p+1 }\G_-
\nn\\
&=&\frac{-\kb}{\sqrt{-det(\CG+\CF)}}[\sum_{\l=0}\{\frac{\8\bslPi^{2\l+1}}
{(2\l+1)!}+\frac{\8\bslPi^{2\l}}{(2\l)!}(\G^\uz D\rho)\}
(1-\bPi'_z  D\rho)e^{\CF'}]_{p+1 }\G_-+...
\nn \\
&=& \frac{-\kb}{\Lam\sqrt{-det(\CG'+\CF')}}[\G^\uz~\sum_{\l=0}
\frac{(\8\bslPi'+\bPi'_z\G^\uz)^
{2\l}}{(2\l)!} D\rho
e^{\CF'}]_{p+1 }\G_-
\nn\\
&=& \frac{-\kb_2}{\sqrt{-det(\CG'+\CF')}}\G^\uz[\sum_{\l=0}
\{\frac{(\slPi')^{2\l}}{(2\l)!}
\}
e^{\CF'}]_{p }\G_-
\nn\\
&=& \frac{-a_1~\kb_1'}{\sqrt{-det(\CG'+\CF')}}
[\sum_{\l=0}\{\frac{(\slPi')^{2\l}}{(2\l)!}\}
e^{\CF'}]_{p }\G_-.
\label{C13}
\eea
Here $[...]_{p+1}$ means p+1 form coefficient of $[...]$ 
, the coefficient of $d\s^0d\s^1...d\s^p$ after taking the pullback.
In the last second line p+1 form coefficient is replaced with
p form coefficient
(the coefficient of $d\s^0d\s^1...d\s^{p-1}$) by dropping $~d\rho~$.
We have also used the relation of DBI term
\bea
{\sqrt{-det(\CG+\CF)}}~=~\Lam~{\sqrt{-det(\CG'+\CF')}}.
\eea

Analogously for the second term in \bref{dtd1},
\bea
-\kb(\gam^{(p)})\G^\uz\G_-
&=&\frac{-\kb}{\sqrt{-det(\CG+\CF)}}[\CS_A(\slPi)e^\CF]_{p+1}
\G^\uz\G_-
\nn\\&=&
\frac{-\kb}{\sqrt{-det(\CG+\CF)}}[\sum_{\l=0}\frac{\slPi^{2\l+1}}
{(2\l+1)!}e^\CF]_{p+1 }\G^\uz(-1)^{\l+1}\G_-
\nn\\&=&
\frac{-\kb}{\sqrt{-det(\CG+\CF)}}[\sum_{\l=0}(-1)^{\l+1}
\frac{(\8\bslPi+\G^\uz D\rho)^{2\l+1}}
{(2\l+1)!}(1+ D\rho\bPi'_z)e^{\CF'}~\G^\uz]_{p+1 }\G_-
\nn\\&=&
\frac{-\kb_1}{\sqrt{-det(\CG'+\CF')}}[\sum_{\l=0}(-1)^{\l+1}
\{\frac{(\slPi')^{2\l}}{(2\l)!}\}e^{\CF'}]_{p }\G_-
\nn\\
&=&\frac{-a_2\kb_2'}{\sqrt{-det(\CG'+\CF')}}
[\sum_{\l=0}(-1)^{\l+1}\frac{(\slPi')^{2\l}}{(2\l)!}
e^{\CF'}]_{p }\G_-.
\label{C15}
\eea
Thus, joining the partial results
\bea
\D\Ttb_1'&=&\kb_1'~-~\frac{a_2}{a_1}~
\frac{\kb_2'}{\sqrt{-det(\CG'+\CF')}}
[\sum_{\l=0}(-1)^{\l+1}\frac{(\slPi')^{2\l}}{(2\l)!}
e^{\CF'}]_{p }\G_-,
\\
\D\Ttb_2'&=&\kb_2'~-~\frac{a_1}{a_2}~\frac{\kb_1'}{\sqrt{-det(\CG'+\CF')}}
[\sum_{\l=0}\{\frac{(\slPi')^{2\l}}{(2\l)!}\}
e^{\CF'}]_{p }\G_-.
\eea
Since $a_1^2=a_2^2=-a_1a_2=1$ we get $a_1/a_2=-1$.
We can express them by using $\tau$ matrices as
\bea
\D\Ttb'&=&\kb'~+~\frac{\kb'}{\sqrt{-det(\CG'+\CF')}}
[\sum_{\l=0}(\tau_3)^{\l+1}~\frac{(\8\slPi')^{2\l}}{(2\l)!}
e^{\CF'}]_{p}~\tau_1~\G_-
\label{C18}
\eea

Thus we have shown that the kappa symmetry transformation of IIA
spinor $\delta_\kappa \overline{\7\theta}$ is mapped to that of IIB spinor
$\delta_{\kappa'} \overline{\7\theta'}$ under $\CT_{\parallel}$,
\bea
\D\Ttb'&=&\kb'(1~-~\gam^{'(p-1)}),
\label{deltatbf}
\\
\gam^{'(p-1)}&=&\frac{-1}{\sqrt{-det(G'+\CF')}}
[\CC_B(\8\slPi')~\tau_1~e^{\CF'}]_{p}.
\eea

\end{document}